\documentclass[12pt,journal,onecolumn]{IEEEtran}
\ifCLASSINFOpdf
\else
\fi
\usepackage{amsmath}
\usepackage{amssymb}
\usepackage{amsthm}
\newtheorem{theorem}{Theorem}
\newtheorem{proposition}[theorem]{Proposition}
\newtheorem{assumption}[theorem]{Assumption}
\newtheorem{corollary}[theorem]{Corollary}
\newtheorem{lemma}[theorem]{Lemma}

\usepackage{mathtools}
\DeclareMathOperator{\Hessian}{Hess}

\usepackage{tikz}
\usetikzlibrary{arrows}
\usepackage{pgfplots}

\usepackage{float}
\usepackage{cuted}

\pgfplotsset{compat=1.13}

\newcommand\Mycomb[2][^n]{\prescript{#1\mkern-0.5mu}{}C_{#2}}

\newcommand{\boldkappa}{\boldsymbol{\kappa}}
\newcommand{\boldeta}{\boldsymbol{\eta}}
\newcommand{\boldbeta}{\boldsymbol{\beta}}
\newcommand{\boldtau}{\boldsymbol{\tau}}
\newcommand{\boldY}{\textbf{Y}}

\DeclarePairedDelimiter\floor{\lfloor}{\rfloor}

\begin{document}
\setlength{\abovedisplayskip}{3pt}
\setlength{\belowdisplayskip}{3pt}
%
\title{Learning to detect an oddball target with observations from an exponential family}
%
%
%

\author{Gayathri R. Prabhu, Srikrishna Bhashyam, Aditya Gopalan, Rajesh Sundaresan
\thanks{This work was supported by the Science and Engineering Research Board, Department of Science and Technology [grant no. EMR/2016/002503]. The authors acknowledge fruitful discussions with Aditya O. Deshmukh.}
\thanks{G. R. Prabhu and S. Bhashyam are with the Department of Electrical Engineering, IIT Madras, Chennai 600036, India.}
\thanks{A. Gopalan is with the Department of Electrical Communication Engineering and R. Sundaresan is with the Department of Electrical Communication Engineering and the Robert Bosch Centre for Cyber-Physical Systems, at the Indian Institute of Science, Bangalore 560012, India.}
} 
\maketitle

\begin{abstract}
The problem of detecting an odd arm from a set of $K$ arms of a multi-armed bandit, with fixed confidence, is studied in a sequential decision-making scenario. Each arm's signal follows a distribution from a vector exponential family. All arms have the same parameters except the odd arm. The actual parameters of the odd and non-odd arms are unknown to the decision maker. Further, the decision maker incurs a cost for switching from one arm to another. This is a sequential decision making problem where the decision maker gets only a limited view of the true state of nature at each stage, but can control his view by choosing the arm to observe at each stage. Of interest are policies that satisfy a given constraint on the probability of false detection. An information-theoretic lower bound on the total cost (expected time for a reliable decision plus total switching cost) is first identified, and a variation on a sequential policy based on the generalised likelihood ratio statistic is then studied. Thanks to the vector exponential family assumption, the signal processing in this policy at each stage turns out to be very simple, in that the associated conjugate prior enables easy updates of the posterior distribution of the model parameters. The policy, with a suitable threshold, is shown to satisfy the given constraint on the probability of false detection. Further, the proposed policy is asymptotically optimal in terms of the total cost among all policies that satisfy the constraint on the probability of false detection.
\end{abstract}


%
\IEEEpeerreviewmaketitle

\section{Introduction}

We consider the problem of detecting an odd arm from a set of $K$ arms of a multi-armed bandit under a \textit{fixed confidence} setting, i.e., with a constraint on the probability of false detection. Each arm follows a distribution from the vector exponential family parameterised by the natural vector parameter $\boldeta$. As the name suggests, all arms except the ``odd" one have the same parameter. The actual parameters of the odd and non-odd arms are unknown. At each successive stage or round, the decision maker chooses exactly one among the $K$ arms for observation. The decision maker therefore has only a limited view of the true state of nature at each stage. But the decision maker can control his view by choosing the arm to observe. The decision maker also incurs a cost whenever he switches from one arm to another. The goal is to minimise the overall cost of expected time for a reliable decision plus total switching cost, subject to a constraint on the probability of false detection. The above serves as a model of how one acquires data during a search task \cite{vaidhiyan17}.

We can model the above problem as a sequential hypothesis testing problem with control \cite{chernoff} and unknown distributions \cite{ref:albert1961sequential} or parameters \cite{vaidhiyan15}. The control here is in the choice of arm for observation at each stage which is determined by the sampling strategy of the policy.

A related problem studied extensively by the machine learning community is that of best arm identification in multi-armed bandits. Garivier et al. \cite{garivier16} have characterised the complexity of best arm identification in one-parameter bandit problems in the fixed confidence setting. Kaufmann et al. \cite{kaufmann16} have discussed the case of identifying $m$ best arms in a stochastic multi-armed bandit model for both fixed confidence and fixed budget settings. In \cite{vaidhiyan17}, the authors have considered the odd arm identification problem with switching costs, but the statistics of the observations were assumed to be known and Poisson-distributed. In \cite{vaidhiyan15}, the authors have considered a learning setting where the parameters of the Poisson distribution were not known but the switching costs were not taken into account. This work provides a significant generalisation of the results in \cite{vaidhiyan15} to the case of a general vector exponential family. This work also analyzes the effect of switching cost on search complexity in the presence of learning, thereby extending the results in \cite{vaidhiyan17} where the parameters were assumed known. For connections to, and limitations of, the works of Chernoff \cite{chernoff} and Albert \cite{ref:albert1961sequential}, see \cite[Sec. I-A]{vaidhiyan15}.


Our interest in the exponential family is for three reasons.
\begin{itemize}
\item It unifies most of the widely used statistical models such as Normal, Binomial, Poisson, and Gamma distributions.
\item The generalisation forces us to rely on, and therefore bring out, the key properties of the exponential family that make the analysis tractable. These include the usefulness of the convex conjugate (or convex dual) of the log partition function, the existence of easily amenable formulae for relative entropy, and the usefulness of the conjugate prior in the analysis.
\item The existence of conjugate priors enables extremely easy posterior updates. This is of great value in practice.
\end{itemize}

We use the results from \cite{kaufmann16} to obtain an information-theoretic lower bound on conditional expected total cost for any policy that satisfies the constraint on probability of false detection, say $\alpha$. The lower bound suggests that the conditional complexity is asymptotically proportional to $\log (1/\alpha)$.

A commonly used test in such problems with unknown parameters is the generalised likelihood ratio test (GLRT) \cite{Poor}. In our case, taking a cue from \cite{vaidhiyan15}, we use a modified GLRT approach where the numerator of the statistic is replaced by an averaged likelihood function. The average is computed with respect to an artificial prior on the unknown parameters. The modified GLRT approach allows us to use a time invariant and a simple threshold policy that meets the constraint on probability of false detection. We show that the sampling strategy of the proposed policy converges to the one suggested by the lower bound as the target probability of false detection $\alpha$ goes down to zero. We also show that, asymptotically, the total cost scales as $\log\left(1/\alpha\right)/D^*$, where $D^*$, a relative entropy based constant, is the optimal scaling factor as suggested by the lower bound.

\subsection{Our contributions}
Our main contributions are the following.
\begin{itemize}
\item We provide a significant generalisation of the odd arm identification problem in \cite{vaidhiyan15}, which dealt with the special case of Poisson observations, to the case of general vector exponential family observations.
\item We modify the policy in \cite{vaidhiyan15} to incorporate switching costs based on the idea of slowed switching in \cite{vaidhiyan17}, \cite{vaidhiyan15_costs} and \cite{krishnaswamy2017augmenting}.
\item We show that the proposed policy, which incorporates learning, is asymptotically optimal even with switching costs; the growth rate of the total cost, as the probability of false detection and the switching parameter are driven to zero, is the same as that without switching costs.
\item  We provide a method to verify an assumption that each arm is sampled at a nontrivial rate. Our rather general approach here, compared to \cite{vaidhiyan15}, provides a simple proof of such a result for Poisson observations. See Appendix \ref{subsubsec:Poisson-distribution}.
\end{itemize}


\subsection{Overview of the proposed policy}
\label{subsec:overview-proposed-policy}

The basic idea of the policy dates back to Chernoff's \textit{Procedure A} \cite{chernoff}. In this work, as indicated above, we modify the generalised likelihood ratio (GLR) function by replacing the maximum likelihood function in the numerator by an average likelihood function. This helps ensure that the policy satisfies the constraint on probability of false detection. We use a time-invariant threshold based on probability of false detection for the policy. Each arm is tested against its nearest alternative by considering the modified GLR function.

At each stage, we choose the arm with the largest GLR statistic. If the statistic exceeds the threshold, we declare the current arm as the odd one and stop further sampling. Else, we decide randomly, based on a coin toss, whether to sample the current arm or choose another one according to the policy's sampling strategy. The bias of the coin determines the speed of switching thereby providing a control on the switching cost. The threshold depends only on the tolerable probability of false detection and the number of arms; it is not time-varying.

Under the vector exponential family assumption, the information processing at each stage is extremely simple. The decision maker maintains the parameters of the associated conjugate priors, corresponding to the posterior distributions of the model parameters, via very simple update rules.

\section{Preliminaries and overview of main result}
In this section we discuss formulae associated with the exponential family that will help in our analysis. We then discuss the model studied and explain the costs under consideration. We end the section with an informal preview of the main result.

\subsection{Exponential family basics}

A probability distribution is a member of a vector exponential family if its probability density function (or probability  mass function) can be written as
\begin{equation}\label{gen_exp_family}
{f\left(x|\boldeta\right) = h\left(x\right)\exp\left(\boldeta^T\textbf{T}(x)-A\left(\boldeta\right)\right)} \quad \forall x,
\end{equation}
where $\boldeta$ is the vector parameter of the family, {{$\boldeta \in \Psi \subset \mathbb{R}^d$ for some $d>0$ {(or $\Psi$ is some open convex subset of $\mathbb{R}^d$)}}}, $\textbf{T}(x) \in \mathbb R^d$ is the sufficient statistic for the family, and $A\left(\boldeta\right)$ is the log partition function given by
\begin{equation*}
A\left(\boldeta\right) = \log \int_{\mathbb R^d}  h\left(x\right)\exp\left(\boldeta^T\textbf{T}(x)\right) dx.
\end{equation*}

The expression in (\ref{gen_exp_family}) gives the {\em canonical} parameterisation of the exponential family. {{We restrict ourselves to minimal representations \cite[p.~40]{wainwright2008graphical} which enables us to represent the distributions in the family using the {\em expectation} parameter defined as}}
\begin{equation}\label{exp_parameter}
{\boldkappa(\boldeta) := } E_{\boldeta}[\textbf{T}(x)]= \nabla_{\boldeta} A\left(\boldeta\right)
\end{equation}

whenever $A{(\cdot)}$ is {continuously} differentiable. The following example will be good to keep in mind.


\textbf{Example (Poisson family):}\label{ex:Poisson}
For the Poisson distribution with alphabet $\mathbb{Z}_+$, we have the probability mass function

\begin{eqnarray}
p\left(x|\lambda\right) &=& \frac{e^{-\lambda}}{x!}\lambda^x \nonumber
			 = \frac{1}{x!} \exp\{x\log \lambda -\lambda\}, \nonumber
\end{eqnarray}

where $\boldeta = \log\lambda$, $\textbf{T}(x) = x$, $A\left(\boldeta\right) = \lambda = e^{\boldeta}$, $h\left(x\right) = \frac{1}{x!}$ and {the expectation parameter is $\boldkappa(\boldeta) = A'(\boldeta) = e^{\boldeta} = \lambda$}. $\hfill \IEEEQEDopen$

\vspace*{.1in}

We now continue with the some additional observations on exponential families. Let us view $A(\boldeta)$ as a function of the parameter $\boldeta$. The mapping $\boldeta \mapsto A(\boldeta)$ is {{strictly convex}}, a fact that can be easily verified via H\"{o}lder inequality. {{The strictness comes from the minimality of the representation.}} Its convex conjugate evaluated at an arbitrary $\boldkappa$ and denoted $F(\boldkappa)$ is given by

\begin{equation}\label{conv_conj_sup}
F\left(\boldkappa\right) := \sup\limits_{\boldeta \in \mathbb{R}^d} \{\boldeta^T \boldkappa-A\left(\boldeta\right)\};
\end{equation}

this is also a convex function. Since $A(\cdot)$ is convex, we obtain that $A(\cdot)$ is recovered as the convex conjugate of $F(\cdot)$, i.e.,

\begin{equation}\label{conv_conj_sup2}
A\left(\boldeta\right) := \sup\limits_{\boldkappa \in \mathbb{R}^d} \{\boldeta^T \boldkappa-F\left(\boldkappa\right)\}.
\end{equation}

We will assume henceforth that {$F(\cdot)$ and $A(\cdot)$} are {{strictly convex and }}twice continuously differentiable at all points where they are finite. {{Optimising (\ref{conv_conj_sup}) over $\boldeta$, recalling that $A(\cdot)$ is strictly convex, we get that the optimising $\boldeta$ is unique and satisfies $\boldkappa = \nabla_{\boldeta} A(\boldeta)$ which is the expectation parameter (\ref{exp_parameter}) evaluated at $\boldeta$.}} Similarly, optimising (\ref{conv_conj_sup2}) over $\boldkappa$, we get an equation similar to (\ref{exp_parameter}), $\boldeta = \nabla_{\boldkappa} F(\boldkappa)$. Thus the optimising $\boldkappa$ and $\boldeta$ are dual to each other and are in one-to-one correspondence. Indeed, we can move from $\boldeta$ to its optimising $\boldkappa$ and from $\boldkappa$ to its optimising $\boldeta$ via
\begin{equation}\label{dual}
\boldkappa \left(\boldeta\right) = \nabla_{\boldeta} A\left(\boldeta\right) \text{ and } \boldeta\left(\boldkappa\right) = \nabla_{\boldkappa} F\left(\boldkappa\right).
\end{equation}

From this one-to-one relation between $\boldeta$ and $\boldkappa$ in (\ref{dual}), we also have

\begin{eqnarray}\label{conv_conj_base}
\begin{array}{c}
F\left(\boldkappa\right) = \boldeta(\boldkappa)^T \boldkappa-A\left(\boldeta(\boldkappa)\right), \\
A\left(\boldeta\right) = \boldeta^T \boldkappa(\boldeta)-F\left(\boldkappa(\boldeta)\right).
\end{array}
\end{eqnarray}

\vspace*{.1in}
When we know that $\boldeta$ and $\boldkappa$ are duals, we simplify the notation in (\ref{conv_conj_base}) to
\begin{eqnarray}\label{conv_conj}
F\left(\boldkappa\right) + A(\boldeta) = \boldeta^T \boldkappa.
\end{eqnarray}

\vspace*{.1in}
That the dual parameter $\boldkappa(\boldeta)$ (respectively, $\boldeta(\boldkappa)$) is involved should be clear from the context (since the supremum in (\ref{conv_conj_sup2}) (respectively, (\ref{conv_conj_sup})) is absent). (See \cite[Section 3.3.2]{Boyd} for these basic properties on convex duals.)

The expressions for KL divergence or relative entropy in terms of the natural parameter and in terms of the expectation parameter (by (\ref{conv_conj})) are
\begin{eqnarray}
\label{eqn:relativeentropy-natural}
D\left(\boldeta_1||\boldeta_2\right) &:=& D\left(f(\cdot|\boldeta_1) || f(\cdot|\boldeta_2) \right) \nonumber \\
							&=& \left(\boldeta_1-\boldeta_2\right)^T \boldkappa_1-A\left(\boldeta_1\right) + A\left(\boldeta_2\right)\\
\label{eqn:relativeentropy-expectation}
				  &=& \left(\boldkappa_2-\boldkappa_1\right)^T\boldeta_2 + F\left(\boldkappa_1\right) - F\left(\boldkappa_2\right).
\end{eqnarray}
Note that we have used the duality relation between $\boldkappa$ and $\boldeta$. The relative entropy $D\left(\boldeta_1||\boldeta_2\right)$ will also be denoted $D\left(\boldkappa_1||\boldkappa_2\right)$ with a minor abuse of notation when we want to make reference to the expectation parameters. These useful formulae will be exploited in later sections.

\subsection{Problem model}
Let $K \geq 3$ be the number of arms available to the decision maker, and let $H$ be the index of the odd arm with $1 \leq H \leq K$. {{Let $\boldeta_1 \in \Psi_1$ and $\boldeta_2 \in \Psi_2$ denote the unknown exponential-family parameter of the odd and non-odd arms, respectively. We assume $\boldeta_1 \neq \boldeta_2$.}} Let the triplet $\psi = \left(i,\boldeta_1,\boldeta_2\right)$ denote the configuration of the arms, where the first component is the index of the odd arm, the second and the third components are the canonical parameters of the odd and non-odd arms, respectively. Let $\mathcal{P}\left(K\right)$ be the set of probability distributions on $\{1,2, \ldots, K\}$.

At any stage, say $n$, given the past observations and actions up to time $n-1$, a policy must choose an action ${\overline{A}_n}$, which is either:
\begin{itemize}
  \item ${\overline{A}_n} = (stop, \delta)$ which is a decision to stop and decide the location of the odd ball as $\delta$, or
  \item ${\overline{A}_n} = (continue, \lambda)$ which is a decision to continue and sample the next arm to pull according to a probability measure on the finite set of arms, $\mathcal{A} = \{1, 2, \ldots, K\}$, returned by a sampling rule $\lambda$.
\end{itemize}
Given a vector of false detection probabilities $\alpha = \left(\alpha_1, \alpha_2,\ldots, \alpha_K\right)$, with each $0 < \alpha_i < 1$, let $\Pi\left(\alpha\right)$ be the set of admissible (desirable) policies that meet the following constraint on the probability of false detection:

\begin{equation}
\Pi\left(\alpha\right) =  \quad \{\pi: P\left(\delta \neq i|\psi = \left(i,\boldeta_1,\boldeta_2\right)\right)\leq \alpha_i, \forall i \text{ and } \forall \psi \text{ such that } \boldeta_1 \neq \boldeta_2\},
\end{equation}
with $\delta$ being the decision made when the algorithm stops. We define the stopping time of the policy as
\begin{equation}
\tau\left(\pi\right) := \inf\{n \geq 1 : {\overline{A}_n} = \left(stop,\cdot \right)\}.
\end{equation}
We also use the notation $||\alpha|| := \max_{i} \alpha_i$.

\subsection{Costs}
The total cost will be the sum of the switching cost and the delay in arriving at a decision as in \cite{vaidhiyan15_costs}. We now make this precise.

\subsubsection{Switching cost}
Let $g\left(a,a'\right)$ denote the cost of switching from {arm} $a$ to {arm} $a'$. We assume
\begin{equation*}
g\left(a,a'\right) \geq 0 \text{ } \forall a, a' \in \mathcal{A} \text{ and } g\left(a,a\right) = 0 \quad \forall a \in \mathcal{A}.
\end{equation*}
The assumption $g(a,a) = 0$ says there is no switching cost if the control does not switch {arms}.
Define $$g_{max} := \max\limits_{a,a' \in \mathcal A} g\left(a,a'\right) < \infty.$$
\subsubsection{Total cost}
For a policy $\pi \in \Pi\left(\alpha\right)$, the total cost $C\left(\pi\right)$ is the sum of stopping time (delay) and net switching cost:
\begin{equation*}
C\left(\pi\right) := \tau\left(\pi\right) + \sum\limits_{l=1}^{\tau\left(\pi\right)-1} g\left(A_l,A_{l+1}\right).
\end{equation*}

\subsection{Informal preview of the main result}
Our main result is to identify the asymptotic growth rate of the cost $\inf_{\pi \in \Pi(\alpha)} C(\pi)$ with respect to $\log (1/||\alpha||)$ as the tolerances for false detection vanish, i.e., $||\alpha|| \rightarrow 0$. We will in particular argue that on account of zero switching cost under no switching and on account of $g_{\max} < \infty$, the switching cost is asymptotically negligible. See Theorem \ref{thm:main-theorem} in Section \ref{sec:main-result} for the precise statement. For an overview of the proposed policy, see the earlier discussion in Section \ref{subsec:overview-proposed-policy}.

\section{The converse (Lower bound on delay)}
\subsection{The lower bound}
The following proposition, available in Albert \cite{ref:albert1961sequential} in a different form, gives an information theoretic lower bound on the expected conditional stopping time for any policy that belongs to $\Pi\left(\alpha\right)$ given the true configuration is $\psi = \left(i,\boldeta_1,\boldeta_2\right)$. We state this converse result here mainly to introduce the relevant quantities for showing achievability.



\begin{proposition}\label{lower_bound}
Fix $\alpha$ with $0 < \alpha_i <1$ for each $i$. Let $\psi = \left(i,\boldeta_1,\boldeta_2\right)$ be the true configuration. For any $\pi \in \Pi\left(\alpha\right)$, we have

\begin{equation}
E\left[\tau|\psi\right] \geq \frac{d_b\left(||\alpha||,1-||\alpha||\right)}{D^*\left(i,\boldeta_1,\boldeta_2\right)}
\end{equation}
where $d_b\left(||\alpha||,1-||\alpha||\right)$ is the binary relative entropy function defined as
{\begin{eqnarray*}
d_b\left(u,1-u\right) :=u\log\left(\frac{u}{1-u}\right)+\left(1-u\right)\log\left(\frac{1-u}{u}\right), \quad u \in [0,1],
\end{eqnarray*}
}
and $D^*\left(i,\boldeta_1,\boldeta_2\right)$ is defined as

\begin{equation} \label{D*}
D^*\left(i,\boldeta_1,\boldeta_2\right) = \max\limits_{\lambda \in \mathcal{P}\left(K\right)} \min\limits_{\boldeta'_1,\boldeta'_2, j \neq i} [\lambda\left(i\right)D\left(\boldeta_1||\boldeta'_2\right)+ \lambda\left(j\right)D\left(\boldeta_2||\boldeta'_1\right)+\left(1-\lambda\left(i\right)-\lambda\left(j\right)\right)D\left(\boldeta_2||\boldeta'_2\right)],
\end{equation}
where $D\left(x||y\right)$ is the relative entropy  (\ref{eqn:relativeentropy-natural}) between two members of the exponential family with natural parameters $x$ and $y$.
\end{proposition}


As the probability of false detection constraint $||\alpha|| \rightarrow 0$, we have $d_b\left(||\alpha||,1-||\alpha||\right)/\log\left(||\alpha||\right)\rightarrow -1$. Hence, we get that the conditional expected stopping time of the optimal policy scales at least as $-\log\left(||\alpha||\right)/D^*\left(i,\boldeta_1,\boldeta_2\right)$. The quantity $D^*\left(i,\boldeta_1,\boldeta_2\right)$ thus characterises the {``complexity''} of the learning problem {at $(i,\boldeta_1,\boldeta_2)$}. A proof of the result may be found in \cite[Prop.~1, p.~4]{vaidhiyan15}.

\begin{corollary}
\label{cor:lowerbound}
We have
\begin{equation}
E[C\left(\pi\right)|\psi] \geq \frac{d_b\left(||\alpha||,1-||\alpha||\right)}{D^*\left(i,\boldeta_1,\boldeta_2\right)}.
\end{equation}
\end{corollary}

\begin{IEEEproof}
With the switching costs added, we have $C\left(\pi\right) \geq \tau\left(\pi\right)$, and the corollary follows from Proposition \ref{lower_bound}.
\end{IEEEproof}

\vspace*{.1in}

We will later show in Theorem \ref{thm:main-theorem} of Section \ref{sec:main-result} that this lower bound is asymptotically tight.

\subsection{A closer look at the problem {complexity} $D^*(i,\boldeta_1,\boldeta_2)$}
Define $\lambda^*\left(i,\boldeta_1,\boldeta_2\right)$ as the $\lambda \in \mathcal{P}\left(K\right)$ that maximises (\ref{D*}). We now study $D^*\left(i,\boldeta_1,\boldeta_2\right)$ and \\ $\lambda^*\left(i,\boldeta_1,\boldeta_2\right)$.

\begin{proposition}\label{optimization}
Fix $K \geq 3$. Let $\psi=\left(i,\boldeta_1,\boldeta_2\right)$ be the true configuration. The quantity in (\ref{D*}) can be expressed as

\begin{equation}\label{eqn:D-opt}
D^*\left(i,\boldeta_1,\boldeta_2\right) =  \max\limits_{0\leq \lambda\left(i\right)\leq 1}\Big[\lambda\left(i\right)D\left(\boldeta_1||\tilde{\boldeta}\right)+\left(1-\lambda\left(i\right)\right)\frac{K-2}{K-1}D\left(\boldeta_2||\tilde{\boldeta}\right)\Big],
\end{equation}
where
\begin{equation}
\label{eqn:etatilde}
\tilde{\boldeta} = {\boldeta}\left(\tilde{\boldkappa}\right),
\end{equation}
{with $\boldeta(\cdot)$ being the function in (\ref{dual})} and
\begin{equation}
\label{eqn:kappatilde}
\tilde{\boldkappa} = \frac{\lambda\left(i\right)\boldkappa_1+\left(1-\lambda\left(i\right)\right)\frac{K-2}{K-1}\boldkappa_2}{\lambda\left(i\right)+\left(1-\lambda\left(i\right)\right)\frac{K-2}{K-1}}.
\end{equation}
Also, $\lambda^*\left(i,\boldeta_1,\boldeta_2\right)$ is of the form
\begin{equation}
\label{eqn:lambda-opt}
\lambda^*\left(i,\boldeta_1,\boldeta_2\right)\left(j\right) = \begin{cases}				\lambda^*\left(i,\boldeta_1,\boldeta_2\right)\left(i\right), & \text{if } j=i \\
									\frac{1-\lambda^*\left(i,\boldeta_1,\boldeta_2\right)\left(i\right)}{K-1}, & \text{if } j\neq i.
									\end{cases}.
\end{equation}
\end{proposition}


\begin{IEEEproof}
Since $\boldeta_1'$ appears only in the middle term in the right-hand side of (\ref{D*}),  it can be minimised by choosing $\boldeta_1' = \boldeta_2$, which makes the term $\lambda\left(j\right)D\left(\boldeta_2||\boldeta'_1\right)$ zero. We therefore have

\begin{eqnarray}
D^*\left(i, \boldeta_1, \boldeta_2\right) &=& \max\limits_{\lambda \in \mathcal{P}\left(K\right)} \min\limits_{\boldeta'_2, j \neq i} [\lambda\left(i\right)D\left(\boldeta_1||\boldeta_2'\right)+ \left(1-\lambda\left(i\right)-\lambda\left(j\right)\right)D\left(\boldeta_2||\boldeta_2'\right)]\label{opti_eta2} \\
					   &=& \max\limits_{0\leq \lambda\left(i\right)\leq 1} \min\limits_{\boldeta_2'} [\lambda\left(i\right)D\left(\boldeta_1||\boldeta_2'\right)+  \left(1-\lambda\left(i\right)\right)\frac{K-2}{K-1}D\left(\boldeta_2||\boldeta_2'\right)]. \label{opti__lambda}
\end{eqnarray}
Equation (\ref{opti__lambda}) follows from the fact that the $\lambda$ that maximises (\ref{opti_eta2}) will have equal mass on all locations other than $i$, i.e.,
\begin{equation*}
\lambda^*\left(j\right) = \frac{1-\lambda^*\left(i\right)}{K-1}, \forall j \neq i.
\end{equation*}
This establishes (\ref{eqn:lambda-opt}).

For a fixed $\lambda\left(i\right)$, to find the $\boldeta_2'$ that minimises the expression in (\ref{opti_eta2}), on account of the strict convexity of the mappings $\boldeta'_2 \mapsto D(\boldeta_1||\boldeta'_2)$ and $\boldeta'_2 \mapsto D(\boldeta_2||\boldeta'_2)$, we take its gradient with respect to $\boldeta_2'$ and equate it to zero. We therefore obtain

\begin{equation}
\label{eqn:lambda-firstorder}
\lambda\left(i\right)\nabla_{\boldeta_2'}D\left(\boldeta_1||\boldeta_2'\right)+\left(1-\lambda\left(i\right)\right)\frac{K-2}{K-1}\nabla_{\boldeta_2'}D\left(\boldeta_2||\boldeta_2'\right) = 0.
\end{equation}

It is easy to see that $\nabla_{\boldeta_2}D\left(\boldeta_1||\boldeta_2\right) = \boldkappa_2-\boldkappa_1$. Plugging this into (\ref{eqn:lambda-firstorder}), we get $\boldkappa'_2$ as

\begin{equation}\label{kappa}
\tilde{\boldkappa} = \boldkappa'_2 = \frac{\lambda\left(i\right)\boldkappa_1+\left(1-\lambda\left(i\right)\right)\frac{K-2}{K-1}\boldkappa_2}{\lambda\left(i\right)+\left(1-\lambda\left(i\right)\right)\frac{K-2}{K-1}}
\end{equation}
and the corresponding $\boldeta$ is obtained using (\ref{dual}) as $\tilde{\boldeta} = {\boldeta}\left(\tilde{\boldkappa}\right).$ This completes the proof of the proposition.
\end{IEEEproof}

\subsection{Nontrivial sampling of all actions}
The quantity $\lambda^*\left(i,\boldeta_1,\boldeta_2\right)$, as a distribution over arms, can be interpreted as a randomised sampling strategy that ``guards" $(i,\boldeta_1,\boldeta_2)$ against its nearest alternative. Heuristically, one would expect an optimal policy's sampling distribution, over the arms, to approach the distribution $\lambda^*\left(i,\boldeta_1,\boldeta_2\right)$ as $||\alpha|| \rightarrow 0$. A closed form expression for $\lambda^*\left(i,\boldeta_1,\boldeta_2\right)$ is not yet available.
\vspace*{.1in}

\begin{assumption}\label{sampling}
Fix $K\geq 3$. Let $\lambda^*$ maximise (\ref{D*}). There exists a constant $c_K \in \left(0,1\right)$, independent of $\left(k,\boldeta_1,\boldeta_2\right)$ but dependent on $K$, such that
\begin{equation*}
\lambda^*\left(k,\boldeta_1,\boldeta_2\right)\left(j\right) \geq c_K >0
\end{equation*}
for all $j \in {1,2, \ldots, K}$ and for all $\left(k,\boldeta_1,\boldeta_2\right)$ such that $\boldeta_1 \neq \boldeta_2$.
\end{assumption}

In Appendix \ref{sampling_proof}, we show that the assumption holds true for a wide range of members from the exponential family. Assumption \ref{sampling} suggests that a policy based on $\lambda^*(i,\boldeta_1,\boldeta_2)$ samples each arm at least $c_K$ fraction of time independent of the ground truth. As we will see, this will ensure consistency of the estimated expectation parameters.

\section{A Sluggish and Modified GLRT}
In this section, we discuss the policy that achieves the lower bound in Proposition 1 as the constraint on probability of false detection is driven to zero. This algorithm is a modification of the policy $\pi_M$ discussed in \cite{vaidhiyan15} to incorporate the switching cost. A similar strategy was used in \cite{vaidhiyan17}, \cite{vaidhiyan15_costs} and \cite{krishnaswamy2017augmenting}.

\subsection{Notations}
Let $N_j^n$ denote the number of times the arm $j$ was chosen for observation up to time $n$, i.e.,
\begin{equation}
  \label{eqn:num-samples}
  N_j^n = \sum\limits_{t=1}^n1_{\{A_t=j\}},
\end{equation}
{where $A_t$ is the arm chosen at time $t$. Clearly}  $n=\sum_{j=1}^K N_j^n$. Let $\textbf{Y}_j^n$ denote the sum of sufficient statistic of arm $j$ up to time $n$, i.e.,
\begin{equation}
  \label{eqn:num-points}
  \textbf{Y}_j^n = \sum\limits_{t=1}^{n}\textbf{T(}X_t\textbf{)}1_{\{A_t=j\}}.
\end{equation}
Let $\textbf{Y}^n$ denote the total sum of the sufficient statistic of all arms up to time $n$, i.e., $\textbf{Y}^n = \sum_{j=1}^{K} \textbf{Y}_j^n$.

\subsection{GLR statistic}
{\em Notation}: We will use the letter $f(\cdot)$ to denote all probability density functions. Conditional densities will be denoted $f(\cdot|\cdot)$. The argument(s) will help identify the appropriate random variable(s) whose density (conditional density) is being represented. We also use it to denote {\em likelihoods} and {\em conditional likelihoods} without the normalisation needed to make them probability or conditional probability densities.

Let $f\left(X^n,A^n|\psi = \left(i,\boldeta_1,\boldeta_2\right)\right)$ be the likelihood function of the observations and actions upto time $n$, under the true state of nature $\psi$, i.e.,
\begin{eqnarray}\label{eqn:likelihood}
{f\left(X^n,A^n|\psi=\left(j,\boldeta_1\left(j\right),\boldeta_2\left(j\right)\right)\right)} &=& \left(\prod\limits_{t=1}^nh\left(X_t\right)\right)\exp\Big\{\boldeta_1^T\left(j\right)\textbf{Y}_j^n-N_j^nA\left(\boldeta_1\left(j\right)\right)\Big\} \nonumber \\
			& & \exp\Big\{\boldeta_2^T\left(j\right)\left(\textbf{Y}^n-\textbf{Y}_j^n\right)-\left(n-N_j^n\right)A\left(\boldeta_2\left(j\right)\right)\Big\}.
\end{eqnarray}

When the parameters are unknown, a natural conjugate prior on {$\boldeta_1(j)$ and $\boldeta_2(j)$ enables easy updates of the posterior distribution based on observations. The conjugate prior, also denoted \\ $f\left(\psi=\left(j,\boldeta_1\left(j\right),\boldeta_2\left(j\right)\right)|H=j\right)$, is taken to be a product distribution with each marginal once again coming from an exponential family of the same form and  characterised by the hyper-parameters $\boldtau$ and $n_0$, i.e.,}
\begin{eqnarray}\label{eqn:conjugateprior}
f\left(\psi=\left(j,\boldeta_1\left(j\right),\boldeta_2\left(j\right)\right)|H=j\right) &=& \mathcal{H}\left(\boldtau,n_0\right)\exp\{\boldtau^T\boldeta_1\left(j\right)-n_0A\left(\boldeta_1\left(j\right)\right)\} \nonumber\\
  & & \times \mathcal{H}\left(\boldtau,n_0\right)\exp\{\boldtau^T\boldeta_2\left(j\right)-n_0A\left(\boldeta_2\left(j\right)\right)\}\\
  \label{eqn:alternative-defn-prior}
  &=:& f(\boldeta_1(j)|\boldtau,n_0) \times f(\boldeta_2(j)|\boldtau,n_0),
\end{eqnarray}
{where we would like to reiterate that $f$ is used to denote both the density of $\psi$ given $H=j$ and the density of $\boldeta_1(j)$ and $\boldeta_2(j)$ given the hyper-parameters. The quantity} $\mathcal{H}\left(\boldtau,n_0\right)$ is the normalising factor given by

\begin{equation}\label{normalizing_factor}
\mathcal{H}\left(\boldtau, n_0\right) = \Big[\int \exp\{\boldtau^T\boldeta-n_0A\left(\boldeta\right)\} d\boldeta\Big]^{-1}.
\end{equation}

{In (\ref{eqn:conjugateprior}) and (\ref{eqn:alternative-defn-prior}), the hyper-parameters $\boldtau$ and $n_0$ are identical for both $\boldeta_1(j)$ and $\boldeta_2(j)$ so that the calculations and presentation are simplified. It is easy to extend the analysis for the case of different hyper-parameters.

It follows from (\ref{conv_conj_sup}) and (\ref{dual}) that the maximum likelihood estimates of the odd and non-odd {\em natural or canonical parameters} $\boldeta_1(j)$ and $\boldeta_2(j)$, at time $n$ and under hypothesis $H = j$, are

\begin{equation}
\label{eqn:eta-update}
\hat{\boldeta}_{1}^n\left(j\right) = \boldeta\left(\hat{\boldkappa}_{1}^n\left(j\right)\right) \text{ and } \hat{\boldeta}_{2}^n\left(j\right) = \boldeta\left(\hat{\boldkappa}_{2}^n\left(j\right)\right),
\end{equation}
{{whenever $\hat{\boldeta}(\cdot)$ exists with }} $\hat{\boldkappa}_j^n = \left(\hat{\boldkappa}_{1}^n\left(j\right), \hat{\boldkappa}_{2}^n\left(j\right)\right)$ where
\begin{equation}
\label{eqn:kappa-update}
\hat{\boldkappa}_{1}^n\left(j\right) = \frac{\textbf{Y}_j^n}{N_j^n} \text{ and } \hat{\boldkappa}_{2}^n\left(j\right) = \frac{\textbf{Y}^n-\textbf{Y}_j^n}{n-N_j^n},
\end{equation}
the maximum likelihood estimates of the odd and non-odd {\em expectation parameters} at time $n$ under $H=j$.
{{Consider a sequence $\delta_n \rightarrow 0$. In cases when the maximum likelihood estimates of the canonical parameter does not exist, we choose $\boldeta_1^*(j)$ and $\boldeta_2^*(j)$ as follows
\begin{equation}\label{eqn:ml_choice}
\left|\sup\limits_{\psi:H=j}f(X^n,A^n|\psi)-f(X^n,A^n|\psi=(j,\boldeta_1^*(j),\boldeta_2^*(j)))\right| \leq \delta_n.
\end{equation}}}
It is the extremely simple nature of (\ref{eqn:kappa-update}) (and its translation to the natural or canonical parameter via (\ref{eqn:eta-update})) that provides ease of updating the posterior distribution of $\boldeta_1(j)$ and $\boldeta_2(j)$, given the observations, under $H = j$.
}
{{
We now substitute {(\ref{eqn:eta-update})} into the likelihood function in (\ref{eqn:likelihood}) to get

\begin{eqnarray}
\hat{f}\left(X^n,A^n|H=j\right) &=& f\left(X^n,A^n|\psi= \left(j,\boldeta^*_1(j), \boldeta^*_2(j)\right)\right)\\
					 &=&\left(\prod\limits_{t=1}^nh\left(X_t\right)\right)\exp\left\{\boldeta_1^*(j)^T\left(j\right)\left(\textbf{Y}_j^n\right)-N_j^nA\left(\boldeta^*_1(j)\left(j\right)\right)\right\}\nonumber \\
					 & & \exp\left\{\boldeta_2^*(j)^T\left(\textbf{Y}^n-\textbf{Y}_j^n\right)-\left(n-N_j^n\right)A\left(\boldeta_2^*(j)^T\right)\right\}. \quad \quad \label{MLf}
\end{eqnarray}}}
{Here $\hat{f}$ denotes} the {\em maximum likelihood} of observations and actions till time $n$ under $H=j$. On the other hand, let the {\em averaged likelihood} function at time $n$, averaged according to the artificial prior $f$ in (\ref{eqn:likelihood}) over all configurations $\psi$ with $H=i$, be
\begin{eqnarray}
\tilde{f}\left(X^n,A^n|H=i\right) &:=& \int f\left(X^n,A^n|\psi=\left(i,\boldeta_1\left(i\right),\boldeta_2\left(i\right)\right)\right) f\left(\boldeta_1\left(i\right)|\boldtau,n_0\right)\nonumber \\
      & & \qquad \quad \cdot f\left(\boldeta_2\left(i\right)|\boldtau,n_0\right)d\boldeta_1\left(i\right)d\boldeta_2\left(i\right)\label{eqn:avg}\\
	  &=& \left(\prod\limits_{t=1}^nh\left(X_t\right)\right) \frac{\mathcal{H}\left(\boldtau,n_0\right)}{\mathcal{H}\left(\textbf{Y}_i^n+\boldtau,N_i^n+n_0\right)} \frac{\mathcal{H}\left(\boldtau,n_0\right)}{\mathcal{H}\left(\left(\textbf{Y}^n-\textbf{Y}_i^n\right)+\boldtau, n-N_i^n+n_0\right)}.\label{eqn:avg-final}
\end{eqnarray}
Equality in (\ref{eqn:avg-final}) is obtained by substituting (\ref{eqn:likelihood}) and (\ref{eqn:conjugateprior}) in (\ref{eqn:avg}) and then replacing integral terms using (\ref{normalizing_factor}). We now define the modified GLR as
\begin{eqnarray}
Z_{ij}\left(n\right) &:=& \log \frac{\tilde{f}\left(X^n,A^n|H=i\right)}{\hat{f}\left(X^n,A^n|H=j\right)} \\
		  &=& \log \Big\{\frac{\mathcal{H}\left(\boldtau,n_0\right)}{\mathcal{H}\left(\textbf{Y}_i^n+\boldtau,N_i^n+n_0\right)}\Big\} + \log \Big\{ \frac{\mathcal{H}\left(\boldtau,n_0\right)}{\mathcal{H}\left(\textbf{Y}^n-\textbf{Y}_i^n+\boldtau, n-N_i^n+n_0\right)}\Big\} \nonumber \\
		  & & -\boldeta_1^*(j)^T\textbf{Y}_j^n+N_j^nA\left(\boldeta_1^*(j)\right)-\boldeta_2^*(j)^T\left(\textbf{Y}^n-\textbf{Y}_j^n\right) +\left(n-N_j^n\right)A\left(\boldeta_2^*(j)\right), \label{eqn:GLR_statistic}
\end{eqnarray}
which is arrived at using (\ref{MLf}) and (\ref{eqn:avg-final}). Let
\begin{equation}
Z_i\left(n\right) := \min\limits_{j \neq i} Z_{ij}\left(n\right)
\end{equation}
denote the modified GLR of $i$ against its nearest alternative.
\subsection{The policy $\pi_{SM}\left(L,\gamma\right)$}
{{Fix $L \geq 1$ (a threshold parameter) and $0<\gamma \leq 1$. The policy involves some new variables: $n^a$ is the number of instants when the decision maker {\em actively} samples based on $\lambda^*\left(i^*\left(n\right),{\boldeta^*}_{1}\left({i^*\left(n\right)}\right), {\boldeta^*}_{2}\left({i^*\left(n\right)}\right) \right)$, and $N_i^{n,a}$ is the number of times the arm $i$ is {\em actively} sampled.
We now define the `Sluggish, Modified GLR' policy as follows.
\vspace*{.5cm}
\hrule
\vspace*{.25cm}
\textit{Policy} $\pi_{SM}\left(L,\gamma\right)$: \\
\hrule
\vspace*{.25cm}
Initialize: Sample the first arm $A_1=1$, Set $n^a=1$, $N_1^{n,a}=1$, $N_i^{n,a}=0 \forall i \neq 1$, $N_1^n=1$, $N_i^n=0 \forall i \neq 1$.
At time $n$:
\begin{itemize}
\item Let $i^*\left(n\right)=\arg \max_i Z_i\left(n\right)$, an arm with the largest modified GLR at time $n$. Resolve ties uniformly at random.
\item If $Z_{i^*\left(n\right)} < \log\left(\left(K-1\right)L\right)$ then choose $A_{n+1}$ via:
\begin{itemize}
\item Generate $U_{n+1}$, a Bernoulli($\gamma$) random variable independent of all other random variables.
\item If $U_{n+1}=0$, then $A_{n+1} = A_n$.
\item If $U_{n+1} = 1$, then update $n^a=n^a+1$ and choose $A_{n+1}$ according to \\
$\lambda^*\left(i^*\left(n\right),{\boldeta^*}_{1}\left({i^*\left(n\right)}\right), {\boldeta^*}_{2}\left({i^*\left(n\right)}\right) \right)$.\\
Resolve ties uniformly at random.\\
Update $N_i^{n,a} = N_i^{n,a}+1$, whenever $A_{n+1}=i$.
\end{itemize}
\item If $Z_{i^*\left(n\right)} \geq \log\left(\left(K-1\right)L\right)$ stop and declare $i^*\left(n\right)$ as the odd arm location.
\end{itemize}}}
\hrule
\vspace*{.1in}

As done in \cite{vaidhiyan15}, we also consider two variants of $\pi_{SM}\left(L,\gamma\right)$ which are useful in the analysis.

\begin{enumerate}
\item \textit{Policy} $\pi_{SM}^i\left(L,\gamma\right)$ is like policy $\pi_{SM}\left(L,\gamma\right)$ but stops only at decision $i$, when $Z_i\left(n\right)\geq \\ \log\left(\left(K-1\right)L\right)$.
\item \textit{Policy} $\tilde{\pi}_{SM}$ is also like $\pi_{SM}\left(L,\gamma\right)$ but never stops.
\end{enumerate}

\section{Achievability preliminaries}

The main steps of the analysis in this section will verify that the above policy
\begin{enumerate}
  \item stops in finite time,
  \item belongs to the desired set of policies, and
  \item is asymptotically optimal.
\end{enumerate}
The above will enable us to establish the main result which is reported { in the next section}. Throughout, Assumption \ref{sampling} is taken to be valid.

\subsubsection{Probability of stopping in finite time}
We assert the following.
\begin{proposition}\label{finite_stopping_time}
Fix the threshold parameter $L>1$. Policy $\pi_{SM}\left(L,\gamma\right)$ stops in finite time with probability 1, that is, $P\left(\tau\left(\pi_{SM}\left(L,\gamma\right)\right)<\infty\right) = 1$.
\end{proposition}


\begin{IEEEproof}
To prove this, we show that when the odd arm has the index $H=i$, the test statistic $Z_i\left(n\right)$ has a positive drift and crosses the threshold $\log\left(\left(K-1\right)L\right)$ in finite time, almost surely. See Appendix \ref{finite_stopping_time_proof}.
\end{IEEEproof}

\subsubsection{Probability of false detection}
We next assert that under a suitable choice of $L$, the proposed policy satisfies the constraint on probability of false detection.


\begin{proposition}
\label{prop:admissible}
Fix $\alpha = \left(\alpha_1, \alpha_2, \ldots, \alpha_K\right)$. Let $L=1/\min_k \alpha_k$. We then have $\pi_{SM}\left(L,\gamma\right) \in \Pi\left(\alpha\right)$.
\end{proposition}


\begin{IEEEproof}
This proof uses elementary change of measure properties, Proposition \ref{finite_stopping_time}, and the result that the policy stops and makes the decision when the statistic $Z_{i^*(n)}$ exceeds the threshold. The proof is identical to that of \cite[Prop.5,~p.8]{vaidhiyan15}.
\end{IEEEproof}

\subsubsection{Asymptotic optimality of the total cost}
The following is an assertion on the drift for the statistic associated with the true odd arm location.

\vspace*{.1in}

\begin{proposition}\label{correct_drift}
Consider the non-stopping policy $\tilde{\pi}_{SM}$. Let $\psi=\left(i,\boldeta_1,\boldeta_2\right)$ be the true configuration. Then,
\begin{equation}
\lim\limits_{n \rightarrow \infty} \frac{Z_i\left(n\right)}{n} \geq D^*\left(i,\boldeta_1,\boldeta_2\right) \text{ a.s.}
\end{equation}
\end{proposition}

\begin{IEEEproof}
See Appendix \ref{upper_bound_proof}.
\end{IEEEproof}


\subsubsection{Achievability} With these ingredients, we can now state the main achievability result. This involves a statement on both the stopping time and on the total cost. The proof uses the above three propositions.


\begin{proposition}\label{upper_bound}
Consider the policy $\pi_{SM}\left(L,\gamma\right)$. Let $\psi=\left(i,\boldeta_1,\boldeta_2\right)$ be the true configuration. Then,
\begin{equation}\label{eqn:upper_bound1}
\limsup\limits_{L \rightarrow \infty} \frac{\tau\left(\pi_{SM}\left(L,\gamma\right)\right)}{\log\left(L\right)} \leq \frac{1}{D^*\left(i,\boldeta_1,\boldeta_2\right)} \text{ a.s.,}
\end{equation}
\begin{equation}\label{eqn:upper_bound2}
\limsup\limits_{L \rightarrow \infty} \frac{E [\tau\left(\pi_{SM}\left(L,\gamma\right)\right)|\psi]}{\log\left(L\right)} \leq \frac{1}{D^*\left(i,\boldeta_1,\boldeta_2\right)},
\end{equation}
and further,
\begin{equation}\label{eqn:upper_bound3}
\limsup\limits_{L \rightarrow \infty} \frac{E[C\left(\pi_{SM}\left(L,\gamma\right)\right)|\psi]}{\log\left(L\right)} \leq \frac{1}{D^*\left(i,\boldeta_1,\boldeta_2\right)} + \frac{g_{max}\gamma}{D^*\left(i,\boldeta_1,\boldeta_2\right)}.
\end{equation}
\end{proposition}


\begin{IEEEproof}
See Appendix \ref{upper_bound_proof}.
\end{IEEEproof}


\section{The main result}
\label{sec:main-result}

With all the above, we can now state and prove the main result.


\begin{theorem}
\label{thm:main-theorem}
Consider $K$ arms with configuration $\psi=\left(i,\boldeta_1,\boldeta_2\right)$. Let $\left(\alpha^{\left(n\right)}\right)_{n\geq 1}$ be a sequence of tolerance vectors such that $\lim\limits_{n \rightarrow \infty} ||\alpha^{\left(n\right)}|| = 0$ and for some finite $B$,
\begin{equation}
\label{eqn:max-min-alpha}
\limsup\limits_{n \rightarrow \infty} \frac{||\alpha^{\left(n\right)}||}{\min_k \alpha_k^{\left(n\right)}} \leq B.
\end{equation}
Then, for each $n$, the policy $\pi_{SM}\left(L_n,\gamma\right)$ with $L_n = 1/\min_k \alpha_k^{\left(n\right)}$ belongs to $\Pi\left(\alpha^{\left(n\right)}\right)$. Furthermore,
\begin{eqnarray}
\liminf\limits_{n \rightarrow \infty} \inf\limits_{\pi \in \Pi\left(\alpha^{\left(n\right)}\right)} \frac{E[C\left(\pi\right)|\psi]}{\log\left(L_n\right)} &=& \lim\limits_{\gamma \rightarrow 0} \lim\limits_{n \rightarrow \infty} \frac{E[C\left(\pi_{SM}\left(L_n,\gamma\right)\right)|\psi]}{\log\left(L_n\right)}\\
		    &=& \frac{1}{D^*\left(i,\boldeta_1,\boldeta_2\right)}.
\end{eqnarray}
\end{theorem}


\begin{IEEEproof}
From Proposition \ref{lower_bound} and (\ref{eqn:max-min-alpha}), it is easy to see that for any admissible policy, the expected stopping time (under $\psi$) grows at least as $(\log(L_n))/D^*(i, \boldeta_1, \boldeta_2)$. From Corollary \ref{cor:lowerbound}, the expected cost too grows at least as $(\log(L_n))/D^*(i, \boldeta_1, \boldeta_2)$. From {Proposition \ref{prop:admissible}}, the policy $\pi_{SM}\left(L_n,\gamma\right)$ is admissible {and, from Proposition \ref{upper_bound},} has an asymptotically growing cost of at most $(1 + g_{max} \gamma) (\log L_n) / D^*(i, \boldeta_1,\boldeta_2)$. Taking $\gamma$ arbitrarily close to 0, we see that we can approach the lower bound. This establishes the theorem.
\end{IEEEproof}

\section{Simulation results}

\begin{figure*}
\centering
\begin{minipage}[b]{.4\textwidth}
\begin{tikzpicture}[scale=.7]
\begin{axis}
[xlabel = $\log(L)$, ylabel = Average stopping time, xmin =0, xmax=250,ymin=0, ymax=2500,
xtick={0,50,100,150,200,250},
xticklabels={0,50,100,150,200,250},
ytick={0,500,1000,1500,2000,2500},
yticklabels={0,500,1000,1500,2000,2500},
legend pos=north west
]
\addplot[mark = triangle*,black,thick,mark options={fill=white}] coordinates{
	(0, 121.6)
	(50, 600.8)
	(100, 1084.1)
	(150, 1525.8)
	(200, 1988.5)
	(250, 2395.9)
};
\addlegendentry{$\gamma = 0.1$}

\addplot[mark = square*,black,thick, mark options={fill=white}] coordinates{
	(0, 89.2)
	(50, 545.8)
	(100, 1006.8)
	(150, 1458.2)
	(200, 1916.6)
	(250, 2339.7)
};
\addlegendentry{$\gamma = 0.5$}

\addplot[mark = *,black,thick, mark options={fill=white}] coordinates{
	(0, 86.7)
	(50, 557)
	(100, 1023)
	(150, 1445.9)
	(200, 1870.3)
	(250, 2339.8)
};
\addlegendentry{$\gamma = 1$}

\addplot[black,thick] coordinates{
	(0, 0)
	(50, 432.7)
	(100,865.4 )
	(150,1298.1 )
	(200, 1730.8)
	(250, 2163.5)
};
\addlegendentry{Lower bound}		
\end{axis}
\end{tikzpicture}
\caption{Performance of $\pi_{SM}(\gamma,L)$ for Gaussian distribution with unknown means. $\mu_1 = 0$, $\sigma_1^2=1$, $\mu_2=1$, $\sigma_2^2 = 1$, $K=8$ and $D^*=0.1156$.}
\label{fig_gaussian_um}
\end{minipage}
\quad
\begin{minipage}[b]{.4\textwidth}
\begin{tikzpicture}[scale=.7]
\begin{axis}
[xlabel = $\log(L)$, ylabel = Average stopping time, xmin =0, xmax=250,ymin=0, ymax=700,
xtick={0,50,100,150,200,250},
xticklabels={0,50,100,150,200,250},
ytick={0,100,200,300,400,500,600,700},
yticklabels={0,100,200,300,400,500,600,700},
legend pos=north west
]

\addplot[mark = triangle*,black,thick,mark options={fill=white}] coordinates{
	(0, 55.19)
	(50, 176.01)
	(100,278.6 )
	(150, 411.9)
	(200, 503.5)
	(250, 614.9)
};
\addlegendentry{$\gamma = 0.1$}

\addplot[mark = square*,black,thick, mark options={fill=white}] coordinates{
	(0, 19.84)
	(50, 135.4)
	(100,249.7 )
	(150, 350.4)
	(200, 461.6)
	(250, 565.6)
};
\addlegendentry{$\gamma = 0.5$}

\addplot[mark = *,black,thick, mark options={fill=white}] coordinates{
	(0, 17.8)
	(50, 128.7)
	(100,236.1 )
	(150,347.8 )
	(200,454.6 )
	(250,559.2 )
};
\addlegendentry{$\gamma = 1$}

\addplot[black,thick] coordinates{
	(0, 0)
	(50, 104.01)
	(100, 208.02)
	(150, 312.03)
	(200, 416.04)
	(250, 520.05)
};
\addlegendentry{Lower bound}	
\end{axis}
\end{tikzpicture}
\caption{Performance of $\pi_{SM}(\gamma,L)$ for Gaussian distribution with unknown variances. $\mu_1=0$, $\sigma^2_1 =25$, $\mu_2=0$, $\sigma^2_2 =1$, $K=8$ and $D^*=0.4807$. }
\label{fig_gaussian_uv}
\end{minipage}

\vspace*{.2in}

\begin{minipage}[b]{.4\textwidth}
\begin{tikzpicture}[scale=.7]
\begin{axis}
[xlabel = $\log(L)$, ylabel = Average stopping time, xmin =0, xmax=250,ymin=0, ymax=1200,
xtick={0,50,100,150,200,250},
xticklabels={0,50,100,150,200,250},
ytick={0,100,200,300,400,500,600,700,800,900,1000,1100,1200},
yticklabels={0,100,200,300,400,500,600,700,800,900,1000,1100,1200},
legend pos=north west
]

\addplot[mark = triangle*,black,thick,mark options={fill=white}] coordinates{
	(0, 206.9)
	(50, 380.9)
	(100,602.7 )
	(150, 784.6)
	(200, 974.3)
	(250, 1186.5)
};
\addlegendentry{$\gamma = 0.1$}

\addplot[mark = square*,black,thick, mark options={fill=white}] coordinates{
	(0, 157)
	(50, 317.2)
	(100, 514.2)
	(150, 731.1)
	(200, 910.6)
	(250, 1103.6)
};
\addlegendentry{$\gamma = 0.5$}

\addplot[mark = *,black,thick, mark options={fill=white}] coordinates{
	(0, 149.2)
	(50,328.3 )
	(100,504.2)
	(150,717.3)
	(200, 905.0)
	(250, 1098.2)
};
\addlegendentry{$\gamma = 1$}

\addplot[black,thick] coordinates{
	(0, 0)
	(50, 195.6)
	(100,391.2 )
	(150, 586.8)
	(200, 782.4)
	(250, 977.9)
};
\addlegendentry{Lower bound}	
\end{axis}
\end{tikzpicture}
\caption{Performance of $\pi_{SM}(\gamma,L)$ for Bernoulli distribution. $p_1=0.1$, $p_2=0.8$, $K=8$ and $D^*=0.2556$.}
\label{fig_bernoulli}
\end{minipage}
\quad
\begin{minipage}[b]{.4\textwidth}
\begin{tikzpicture}[scale=.7]
\begin{axis}
[xlabel = $\log(L)$, ylabel = Average stopping time, xmin =0, xmax=250,ymin=0, ymax=1000,
xtick={0,50,100,150,200,250},
xticklabels={0,50,100,150,200,250},
ytick={0,100,200,300,400,500,600,700,800,900,1000},
yticklabels={0,100,200,300,400,500,600,700,800,900,1000},
legend pos=north west
]

\addplot[mark = triangle*,black,thick,mark options={fill=white}] coordinates{
	(0, 240.1)
	(50, 352.4)
	(100, 496.2)
	(150, 658.7)
	(200, 768.6)
	(250, 913.1)
};
\addlegendentry{$\gamma = 0.1$}

\addplot[mark = square*,black,thick, mark options={fill=white}] coordinates{
	(0, 75.8)
	(50, 226.8)
	(100, 373.1)
	(150, 516.2)
	(200,670.6 )
	(250, 818.5)
};
\addlegendentry{$\gamma = 0.5$}

\addplot[mark = *,black,thick, mark options={fill=white}] coordinates{
	(0, 57.65)
	(50, 212.37)
	(100, 357.4)
	(150, 507.36)
	(200, 648.8)
	(250, 795.5)
};
\addlegendentry{$\gamma = 1$}

\addplot[black,thick] coordinates{
	(0, 0)
	(50, 143.05)
	(100, 286.1)
	(150, 429.2)
	(200, 572.2)
	(250, 715.3)
};
\addlegendentry{Lower bound}	
\end{axis}
\end{tikzpicture}
\caption{Performance of $\pi_{SM}(\gamma,L)$ for Vector Gaussian distrbution. $\mu_1 =0$, $\sigma_1^2 = 2$, $\mu_2=4$, $\sigma_2^2=5$, $K=8$ and $D^* = 0.3495$.}
\label{fig_vector_gaussian}
\end{minipage}
\end{figure*}

{In this section we study the performance of the proposed policy $\pi_{SM}(L,\gamma)$  for different values of $L$ and switching parameter $\gamma$ using numerical simulations. Fig. \ref{fig_gaussian_um} - Fig. \ref{fig_vector_gaussian} show the empirical average stopping time of our policy averaged over $100$ independent runs plotted against $\log(L)$ for single parameter Gaussian (unknown mean or unknown variance), Bernoulli, and vector parameter Gaussian (both mean and variance unknown) cases. We also plot the lower bound on expected stopping time as suggested by the Proposition \ref{lower_bound}.

The switching parameter is varied from $\gamma=0.1$, which corresponds to a sluggish implementation, to $\gamma=1$ when the policy switches according to the sampling strategy at each stage. As expected, we can make the following observations from the plots: (1) the slope for the policy in each case (and for each $\gamma$) matches with the slope of the lower bound thereby validating the asymptotic optimality of the policy; (2) with a smaller switching parameter, the policy takes more number of samples to arrive at a decision as compared to larger switching parameters.
}

\section{Summary}

In this work, we discussed a policy to detect an odd arm  from a set of arms with minimum cost under a constraint on the probability of false detection. The arm observations are assumed to be sampled from distributions that belong to general exponential families. The total cost is taken as the sum of (1) delay in arriving at a decision and (2) switching cost. The switching of arms is controlled using a Bernoulli random variable with parameter $\gamma$, which slows down the switching. Slowed switching implies that exploration is not done as quickly as in the case with no switching costs. The stopping time however continues to grow at the same asymptotic rate since the arms are sampled with the correct asymptotic marginal distribution, even though in a sluggish and possibly correlated (e.g., Markovian) way. We then obtained that the growth rate of total cost, as both the probability of false detection and the switching parameter $\gamma$ are driven to zero, is the same as that without switching costs. Crucial to our analysis is the assumption that each arm is sampled a nontrivial fraction of times, no matter what the underlying true state of nature. In Appendix \ref{sampling_proof} we demonstrate how to verify the condition in a few important examples.

\appendices
\section{Assumption \ref{sampling}:
nontrivial sampling of all actions} \label{sampling_proof}
In this section, we show that many common exponential families satisfy Assumption \ref{sampling}.
We begin by re-writing the expression (\ref{eqn:D-opt}) as
\begin{eqnarray}
\lambda^*\left(k,\boldeta_1,\boldeta_2\right)\left(i\right)=\arg \max\limits_{0\leq \lambda \leq 1}\left[\lambda D\left(\boldeta_1||\tilde{\boldeta}\right)+\left(1-\lambda\right)\frac{K-2}{K-1}D\left(\boldeta_2||\tilde{\boldeta}\right)\right]. \quad \quad \label{lambda*1}
\end{eqnarray}
Note that $\tilde{\boldeta}$ depends on $\lambda$ as per (\ref{eqn:etatilde}) and (\ref{eqn:kappatilde}). As a first step, we show that the optimisation problem (\ref{lambda*1}) is concave, and then obtain a bound on the value of $\lambda$ that achieves this maximum. To establish the concavity, we show that the second derivative of the objective function in (\ref{lambda*1}) is nonpositive for all $\lambda$. Define the objective function in (\ref{lambda*1}) as
\begin{equation*}
\Phi\left(\lambda\right) := \lambda D\left(\boldeta_1||\tilde{\boldeta}\right)+\left(1-\lambda\right)\frac{K-2}{K-1}D\left(\boldeta_2||\tilde{\boldeta}\right)
\end{equation*}
where $\tilde{\boldeta}$ is also a function of $\lambda$. Taking derivative, we get
\begin{eqnarray}
\frac{d\Phi}{d\lambda} &=& D\left(\boldeta_1||\tilde{\boldeta}\right) - \frac{K-2}{K-1}D\left(\boldeta_2||\tilde{\boldeta}\right)+ \left[\lambda \nabla_{\tilde{\boldeta}}D\left(\boldeta_1||\tilde{\boldeta}\right)+\left(1-\lambda\right)\frac{K-2}{K-1}\nabla_{\tilde{\boldeta}}D\left(\boldeta_2||\tilde{\boldeta}\right)\right]^T\frac{d\tilde{\boldeta}}{d\lambda}\label{eqn:PhiDerivative}\\
					   &=& D\left(\boldeta_1||\tilde{\boldeta}\right) - \frac{K-2}{K-1}D\left(\boldeta_2||\tilde{\boldeta}\right). \label{Phi:first_deriv}
\end{eqnarray}
Equality in (\ref{Phi:first_deriv}) follows from (\ref{eqn:lambda-firstorder}), which ensures that the term within square brackets in (\ref{eqn:PhiDerivative}) is zero. Differentiating again,
\begin{eqnarray}
\frac{d^2\Phi}{d\lambda^2}
						   &=& \left[\left(\tilde{\boldkappa}-\boldkappa_1\right) - \frac{K-2}{K-1}\left(\tilde{\boldkappa}-\boldkappa_2\right)\right]^T\frac{d\tilde{\boldeta}}{d\lambda} < 0. \quad \label{concavity}
\end{eqnarray}
The equality in (\ref{concavity}) follows from $\nabla_{\boldeta_2} D\left(\boldeta_1||\boldeta_2\right) = \boldkappa_2-\boldkappa_1$, and  the inequality in (\ref{concavity}) is obtained using
\begin{eqnarray}
\frac{d\tilde{\boldeta}}{d\lambda} &=& D_{\tilde{\boldkappa}} \tilde{\boldeta} \cdot \frac{d\tilde{\boldkappa}}{d\lambda} \label{eqn:eta_firstord_1}\\
							&=& \Hessian(F(\tilde{\boldkappa})) \cdot \frac{(-1)}{\lambda+\left(1-\lambda\right)\frac{K-2}{K-1}}\left(\left(\tilde{\boldkappa}-\boldkappa_1\right)-\frac{K-2}{K-1}\left(\tilde{\boldkappa}-\boldkappa_2\right)\right). \label{eqn:eta_firstord_2}
\end{eqnarray}
Equation (\ref{eqn:eta_firstord_1}), where $D_{\tilde{\boldkappa}} \tilde{\boldeta}$ is the matrix $\left(\frac{\partial}{\partial \tilde{\boldkappa}_j} \tilde{\boldeta}_i\right)_{1 \leq i,j \leq d}$, follows from the chain rule for differentiation. From (\ref{dual}), we recognise that $D_{\tilde{\boldkappa}} \tilde{\boldeta} = \Hessian(F(\tilde{\boldkappa}))$, the hessian of the function $F(\boldkappa)$ with respect to $\boldkappa$ evaluated at $\tilde{\boldkappa}$. Using this and a straightforward calculation of the derivative $d\tilde{\boldkappa}/d\lambda$, we get (\ref{eqn:eta_firstord_2}). Substituting (\ref{eqn:eta_firstord_2}) in (\ref{concavity}) and using the fact that the Hessian of {{the strictly convex function $F(\boldkappa)$ is positive definite}}, we obtain the result in (\ref{concavity}).

Since $\Phi\left(\lambda\right)$ is concave in $\lambda$, and since $\Phi\left(0\right) = \Phi\left(1\right) = 0$ and $\Phi'\left(0\right)>0$ and $\Phi'\left(1\right)<0$, maximiser $\lambda^*$ satisfies
\begin{equation}\label{eqn:firstderivative_eta}
D\left(\boldeta_1||\tilde{\boldeta}\right) - \frac{K-2}{K-1}D\left(\boldeta_2||\tilde{\boldeta}\right) = 0.
\end{equation}
We do not know a closed form expression for $\lambda^*$ from (\ref{eqn:firstderivative_eta}). Let $\hat{\lambda}$ denote a parameterisation of $\lambda$ of the form
\begin{equation}
\hat{\lambda} = \frac{\lambda}{\lambda+\left(1-\lambda\right)\frac{K-2}{K-1}}
\end{equation}
so that $\tilde{\boldkappa} = \hat{\lambda}\boldkappa_1 + \left(1-\hat{\lambda}\right)\boldkappa_2$.
We can see that $\hat{\lambda}$ is increasing in $\lambda$. Also, let $\hat{\lambda}^*$ denote the reparameterisation of $\lambda^*$. Hence, to show that $\lambda^*$ is bounded away from $0$ and $1$, it suffices to show that $\hat{\lambda}^*$ is bounded away from $0$ and $1$.

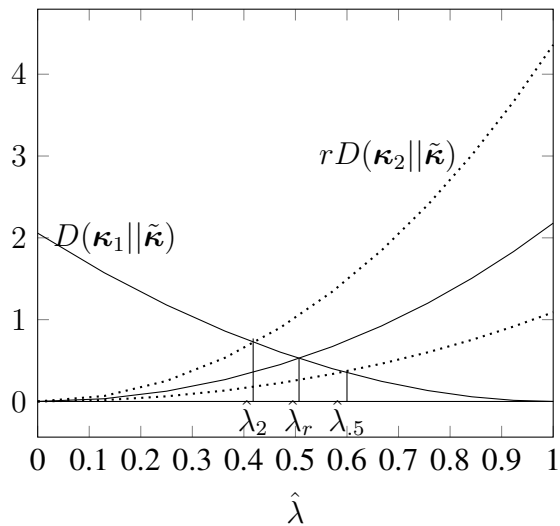
\begin{figure}[t]
\centering
\begin{tikzpicture}[scale=1]
\begin{axis}
[xlabel=$\hat{\lambda}$, xmin=0,xmax=1,
xtick={0,0.1,0.2,0.3,0.4,0.5,0.6,0.7,0.8,0.9,1.0},
xticklabels={0,0.1,0.2,0.3,0.4,0.5,0.6,0.7,0.8,0.9,1}
]
\addplot[black,thick, dotted] coordinates {
	(0 , 0 )
    (.129 , .0167 )
    (.25 , .0628)
    (.3636 , .1337)
    (.4706 , .2255)
    (.5714 , .3354)
    (.6667 , .4610)
    (.7568 , .6006)
    (.8421 , .7527)
    (.9231 , .9160)
    (1  , 1.0897 )
  } ;
\addplot[black, mark=none] coordinates {
	(0 , 0 )
    (.129 , .0333 )
    (.25 , .1256)
    (.3636 , .2673)
    (.4706 , .4509)
    (.5714 , .6707)
    (.6667 , .9221)
    (.7568 , 1.2013)
    (.8421 , 1.5054)
    (.9231 , 1.8321)
    (1  , 2.1794 )
  } ;
\addplot[black, thick, dotted] coordinates {
	(0 , 0 )
    (.129 , .0667 )
    (.25 , .2513)
    (.3636 , .5346)
    (.4706 , .9019)
    (.5714 , 1.3415)
    (.6667 , 1.8442)
    (.7568 , 2.4026)
    (.8421 , 3.0108)
    (.9231 , 3.6642)
    (1  , 4.3588)
  } ;

\addplot[black, mark=none] coordinates {
	(0 , 2.0571 )
    (.129 , 1.5738 )
    (.25 , 1.1793)
    (.3636 , 0.8594)
    (.4706 , 0.6029)
    (.5714 , 0.4010)
    (.6667 , 0.2464)
    (.7568 , 0.1334)
    (.8421 , 0.0572)
    (.9231 , 0.0138)
    (1  ,  0)
  } ;
\node[] at (axis cs: 0.15,2) {$D(\boldkappa_1||\tilde{\boldkappa})$};
\node[] at (axis cs: 0.68,3) {$r D(\boldkappa_2||\tilde{\boldkappa})$};
\addplot [black, mark=none] coordinates {(0,0) (1,0)};
\node[] at (axis cs: 0.418,-0.2) {$\hat{\lambda}_2$};
\addplot [black, mark=none] coordinates {(.418,0) (.418,.77)};
\node[] at (axis cs: 0.507,-0.2) {$\hat{\lambda}_r$};
\addplot [black, mark=none] coordinates {(.507,0) (.507,.52)};
\node[] at (axis cs: 0.6,-0.2) {$\hat{\lambda}_{.5}$};
\addplot [black, mark=none] coordinates {(.6,0) (.6,.37)};
\end{axis}
\end{tikzpicture}
\caption{Geometric interpretation of $\hat{\lambda}^*$.}
\label{fig:interpretation}
\end{figure}

We re-write the expression in (\ref{eqn:firstderivative_eta}) in terms of the expectation parameter $\boldkappa$ for ease of representation and computation.
\begin{equation}\label{eqn:firstderivative_kappa}
D\left(\boldkappa_1||\tilde{\boldkappa}\right) - rD\left(\boldkappa_2||\tilde{\boldkappa}\right) = 0,
\end{equation}
with $r = \frac{K-2}{K-1}$. Fig. \ref{fig:interpretation} gives a geometric interpretation of $\hat{\lambda}^*$. It can be observed that $\hat{\lambda}^* = \hat{\lambda}_r$ in the picture, and this decreases with $r$. Further, we know $0.5 \leq r \leq 2$ which implies $\hat{\lambda}_{2} < \hat{\lambda}^*_r <\hat{\lambda}_{0.5}$. Hence, to show $\hat{\lambda}^*$ is bounded away from $0$ and $1$, it suffices to show that $\hat{\lambda}_{0.5}<1$ and $\hat{\lambda}_2>0$.

Next, we re-write the expression in (\ref{eqn:firstderivative_kappa}) using Taylor's theorem to ease the computations.
\begin{lemma}\label{lem:Zero-HessIntegral}
Recall the expression for relative entropy $D(\boldkappa_1||\boldkappa_2) = F(\boldkappa_1)-F(\boldkappa_2)-\nabla_{\boldkappa} F(\boldkappa_2)^T(\boldkappa_1-\boldkappa_2)$. Then  (\ref{eqn:firstderivative_kappa}) can be rewritten as
\begin{equation}
\int\limits_{\hat{\lambda}}^{1}(1-u)\Delta\boldkappa^T\Hessian{(F(\boldkappa_2 + u\Delta\boldkappa))}\Delta\boldkappa ~du - ~r \int\limits_{0}^{\hat{\lambda}} u\Delta\boldkappa^T\Hessian{(F(\boldkappa_2 + u\Delta\boldkappa))}\Delta\boldkappa ~du  = 0, \label{eqn:firstderiv_integral}
\end{equation}
where $\Delta\boldkappa = \boldkappa_1-\boldkappa_2$.
\end{lemma}
\begin{IEEEproof}
Since $F(\boldkappa)$ is twice differentiable, use of the multivariate Taylor theorem for $F(\boldkappa_1)$ near $\tilde{\boldkappa}$ yields
\begin{eqnarray}
D(\boldkappa_1||\tilde{\boldkappa}) &=& F(\tilde{\boldkappa})+\nabla_{\boldkappa_1}  F(\tilde{\boldkappa})^T(\boldkappa_1-\tilde{\boldkappa})+\sum\limits_{|\boldbeta|=2}R_{\boldbeta}(\boldkappa_1)(\boldkappa_1-\tilde{\boldkappa})^\beta -  F(\tilde{\boldkappa})-\nabla_{\boldkappa_1}  F(\tilde{\boldkappa})^T(\boldkappa_1-\tilde{\boldkappa}) \nonumber\\
									&=&\sum\limits_{|\boldbeta|=2}R_{\boldbeta}(\boldkappa_1)(\boldkappa_1-\tilde{\boldkappa})^{\boldbeta},\label{eqn:TaylorExp}
\end{eqnarray}
where
\begin{equation}\label{eqn:remainder}
R_{\boldbeta}(\boldkappa_1) = \frac{|\boldbeta|}{\boldbeta!}\int\limits_{0}^{1} (1-t)^{|\boldbeta|-1} D^{\boldbeta} F\left(\tilde{\boldkappa}+t(\boldkappa_1-\tilde{\boldkappa})\right) dt.
\end{equation}
We next discuss each term in (\ref{eqn:TaylorExp}) and (\ref{eqn:remainder}) in detail. Since the vector $\boldkappa$ is $d$ dimensional, we have $\Mycomb[d]{1}+\Mycomb[d]{2}$ possible values for the $d \times 1$ vector $\boldbeta$, such as $(2,0,0,\ldots,0), (0,2,0,\ldots,0),(1,1,0,\ldots,0)$, etc., where the elementwise sum denoted $|\boldbeta|$ adds to $2$. Also, we use the standard multi-index notation ${\boldbeta}! = \beta_1! \beta_2! \ldots \beta_d!$, $x^{\boldbeta} = x_1^{\beta_1}x_2^{\beta_2} \ldots x_d^{\beta_d}$ and
\begin{equation*}
D^{\boldbeta} f(x) = \frac{d^{|\boldbeta|}f(x)}{dx_1^{\beta_1} \ldots dx_d^{\beta_d}}.
\end{equation*}
Using these, we can rewrite (\ref{eqn:TaylorExp}) in matrix form as
\begin{equation}\label{eqn:TaylorExp_LHS}
D(\boldkappa_1||\tilde{\boldkappa}) = \int\limits_{0}^{1} (1-t) \Delta{\boldkappa_1}^T \Hessian{(F\left(\tilde{\boldkappa}+t\Delta\boldkappa_1\right))} \Delta\boldkappa_1 dt,
\end{equation}
where $\Delta \boldkappa_1 = \boldkappa_1-\tilde{\boldkappa}$ and $\Hessian{(F)}$ is the Hessian matrix.

We use $\tilde{\boldkappa} = \hat{\lambda}\boldkappa_1 + (1-\hat{\lambda})\boldkappa_2$ to get $\Delta \boldkappa_1 = (1-\hat{\lambda})\left(\boldkappa_1-\boldkappa_2\right)$, change variables suitably in (\ref{eqn:TaylorExp_LHS}), and simplify to obtain

\begin{equation}
D(\boldkappa_1||\tilde{\boldkappa}) = \int\limits_{\hat{\lambda}}^{1}(1-u)\Delta\boldkappa^T \Hessian{(F(\boldkappa_2 + u\Delta\boldkappa))} \Delta\boldkappa du,
\end{equation}
where $\Delta \boldkappa = \boldkappa_1-\Delta\boldkappa_1$. Following similar steps for $D(\boldkappa_2||\tilde{\boldkappa})$ we get the required result in (\ref{eqn:firstderiv_integral}).
\end{IEEEproof}

Hence, to show that $\hat{\lambda}_{0.5}$ is bounded away from $1$, it suffices to show that the following holds: $\exists \hat{\lambda}^*<1$ such that
\begin{equation}
\int\limits_{\hat{\lambda}^*}^{1}(1-u)\Delta\boldkappa^T\Hessian{(F(\boldkappa_2 + u\Delta\boldkappa))}\Delta\boldkappa ~ du - \frac{1}{2}\int\limits_{0}^{\hat{\lambda}^*} u\Delta\boldkappa^T\Hessian{(F(\boldkappa_2 + u\Delta\boldkappa))}\Delta\boldkappa ~du < 0.\label{eqn:main_ineq_.5}
\end{equation}
Similarly, in order to show $\hat{\lambda}_2>0$, it is enough that the following holds: $\exists \hat{\lambda}^*>0$ such that
\begin{equation}
\int\limits_{\hat{\lambda}^*}^{1}(1-u)\Delta\boldkappa^T\Hessian{(F(\boldkappa_2 + u\Delta\boldkappa))}\Delta\boldkappa ~du - 2\int\limits_{0}^{\hat{\lambda}^*} u\Delta\boldkappa^T\Hessian{(F(\boldkappa_2 + u\Delta\boldkappa))}\Delta\boldkappa ~du  > 0.
\label{eqn:main_ineq_2}
\end{equation}
Multiply (\ref{eqn:main_ineq_2}) throughout by $1/2$, change variables $u$ to $1-u$, and swap $\boldkappa_1$ and $\boldkappa_2$ to see that a search for $\hat{\lambda}_2 > 0$ satisfying (\ref{eqn:main_ineq_2}) for arbitrary $\boldkappa_1, \boldkappa_2$ is identical to a search for $\hat{\lambda}_{0.5} < 1$ solving (\ref{eqn:main_ineq_.5}) for arbitrary $\boldkappa_1, \boldkappa_2$. Hence, in the following sections we proceed to verify (\ref{eqn:main_ineq_.5}).

We do not have a complete solution for the inequality in (\ref{eqn:main_ineq_.5}) for the general exponential family. Instead, we show that this condition, and hence Assumption \ref{sampling} holds true for a few single parameter family members. For the vector parameter Gaussian distribution, we check (\ref{eqn:main_ineq_.5}) numerically.

\subsection{Single parameter distributions}
\subsubsection{Poisson distribution}
\label{subsubsec:Poisson-distribution}
Recall the example in \ref{ex:Poisson}. With $\boldkappa = \lambda$, we compute $F(\boldkappa)$ using (\ref{conv_conj}) as
\begin{equation}
F(\boldkappa) = \boldkappa \log \boldkappa - \boldkappa
\end{equation}
and
\begin{equation}
\frac{dF}{d\boldkappa} = \log \boldkappa,  \frac{d^2F}{d\boldkappa^2} = \frac{1}{\boldkappa}.
\end{equation}
Therefore (\ref{eqn:main_ineq_.5}) requires
\begin{equation}\label{eqn: Poisson_inequality}
\int\limits_{\hat{\lambda}}^{1}(1-u)\frac{\left(\Delta\boldkappa\right)^2}{\boldkappa_2+u\Delta\boldkappa} du < \frac{1}{2}\int\limits_{0}^{\hat{\lambda}} u\frac{\left(\Delta\boldkappa\right)^2}{\boldkappa_2+u\Delta\boldkappa} du
\end{equation}
We proceed further by considering two cases.
\paragraph{$\Delta\boldkappa > 0$}
Using the fact that the second derivative is a decreasing function in $u$ and $\Delta \boldkappa$ is independent of $u$, (\ref{eqn: Poisson_inequality}) holds if
\begin{equation*}
\int\limits_{\hat{\lambda}}^{1}\frac{(1-u)}{\boldkappa_2+\hat{\lambda}\Delta\boldkappa} du < \frac{1}{2}\int\limits_{0}^{\hat{\lambda}} \frac{u}{\boldkappa_2+\hat{\lambda}\Delta\boldkappa} du,
\end{equation*}
or if
\begin{equation}\label{eqn:Poisson_final}
\int\limits_{\hat{\lambda}}^{1} (1-u) du < \frac{1}{2}\int\limits_{0}^{\hat{\lambda}} u du.
\end{equation}
On solving (\ref{eqn:Poisson_final}), we get $\hat{\lambda} > 0.59$ suffices for (\ref{eqn: Poisson_inequality}) to hold.

\paragraph{$\Delta\boldkappa < 0$}
For this case, define $d=\boldkappa_2-\boldkappa_1$. Then (\ref{eqn: Poisson_inequality}) can be written as
\begin{equation*}
\int\limits_{\hat{\lambda}}^{1}\frac{1-u}{\boldkappa_2-ud} du < \frac{1}{2}\int\limits_{0}^{\hat{\lambda}} \frac{u}{\boldkappa_2-ud} du.
\end{equation*}
Rewrite this as
\[
\int\limits_{\hat{\lambda}}^{1}\frac{1}{\boldkappa_2}\frac{1-u}{1-ud/\boldkappa_2} du <  \frac{1}{2}\int\limits_{0}^{\hat{\lambda}} \frac{1}{\boldkappa_2}\frac{u}{1-ud/\boldkappa_2} du.
\]
Since $(1 - u)/(1 - ud/\boldkappa_2) \leq 1$ and since $1/(1 - ud/\boldkappa_2) \geq 1$, we get that (\ref{eqn: Poisson_inequality}) holds if
\begin{eqnarray}
\int\limits_{\hat{\lambda}}^{1}\frac{1}{\boldkappa_2} du <  \frac{1}{2}\int\limits_{0}^{\hat{\lambda}} \frac{u}{{\boldkappa_2}} du,
\end{eqnarray}
which holds if $\hat{\lambda}< 0.82$. Choose a $\hat{\lambda}$ that satisfies both constraints in cases (a) and (b).

\subsubsection{Bernoulli distribution}
\begin{equation}
P(x;p) = \exp\left\{\left(x \log\frac{p}{1-p}\right)+\log(1-p)\right\}
\end{equation}
with $\boldeta = \log\frac{p}{1-p}$, $\textbf{T}(x) = x$, $A(\boldeta) = -\log(1-p)$ and $\boldkappa = p$. We then compute
\begin{equation}
F(\boldkappa) = p\log p +(1-p)\log(1-p),
\end{equation}
and
\begin{equation}
\frac{dF}{d\boldkappa} = \log p - \log(1-p), \frac{d^2F}{d\boldkappa^2} = \frac{1}{p(1-p)}.
\end{equation}
Therefore (\ref{eqn:main_ineq_.5}) yields
\begin{eqnarray}\label{eqn:Bernoulli_inequality}
\int\limits_{\hat{\lambda}}^{1} (1-u)\frac{\left(\Delta \boldkappa\right)^2}{\left(\boldkappa_2+u\Delta\boldkappa\right)\left(1-\left(\boldkappa_2+u\Delta\boldkappa\right)\right)}du & < & \frac{1}{2}\int\limits_{0}^{\hat{\lambda}}u \frac{\left(\Delta \boldkappa\right)^2}{\left(\boldkappa_2+u\Delta\boldkappa\right)\left(1-\left(\boldkappa_2+u\Delta\boldkappa\right)\right)} du
\end{eqnarray}

We do not have an analytical solution for a $\hat{\lambda}$ for which (\ref{eqn:Bernoulli_inequality}) is true. Therefore, we numerically check the inequality in Fig. \ref{fig:Bernoulli} by varying $\boldkappa_1$ and $\boldkappa_2$ in $[0,1]$ and for $\hat{\lambda} \in [0,1]$. From the plot, it can be observed that for $\hat{\lambda}>0.75$ the assumption in (\ref{eqn:main_ineq_.5}) holds.
\begin{figure*}
\centering
\begin{minipage}[t]{.4\textwidth}
\begin{tikzpicture}[scale=.7]
\begin{axis}
[xlabel=$\hat{\lambda}$, ylabel=$\max\limits_{\boldkappa_1,\boldkappa_2}\left\{D(\boldkappa_1||\tilde{\boldkappa})-0.5D(\boldkappa_2||\tilde{\boldkappa})\right\}$, xmin=0,xmax=1, ymin=-3,ymax=15,
xtick={0,0.1,0.2,0.3,0.4,0.5,0.6,0.7,0.8,0.9,1.0},
xticklabels={0,0.1,0.2,0.3,0.4,0.5,0.6,0.7,0.8,0.9,1},
ytick={-3,0,3,6,9,12,15},
yticklabels={-3,0,3,6,9,12,15}
]

\addplot [black, mark = *]coordinates {
    (.1 , 13.9454 )
    (.2 , 8.2115)
    (.3 , 6.9492)
    (.4 , 5.5902)
    (.5 , 4.0966)
    (.6 , 2.4050)
    (.7 , 0.3966)
    (.8 , -0.5092)
    (.9 , -0.7900)
    (1  , -1 )
  } ;
\addplot [black, dotted] coordinates {(0,0) (1,0)};
\end{axis}
\end{tikzpicture}
\caption{Variation of $\max\limits_{\boldkappa_1,\boldkappa_2}\left\{D(\boldkappa_1||\tilde{\boldkappa})-0.5D(\boldkappa_2||\tilde{\boldkappa})\right\}$ with $\hat{\lambda}$ for Bernoulli distribution.}
\label{fig:Bernoulli}
\end{minipage}
\quad
\begin{minipage}[t]{.4\textwidth}
\begin{tikzpicture}[scale=.7]
\begin{axis}
[xlabel=$\hat{\lambda}$, ylabel=$\max\limits_{\boldkappa_1,\boldkappa_2}\left\{D(\boldkappa_1||\tilde{\boldkappa})-0.5D(\boldkappa_2||\tilde{\boldkappa})\right\}$, xmin=0,xmax=1,ymin=-3,ymax=15,
xtick={0,0.1,0.2,0.3,0.4,0.5,0.6,0.7,0.8,0.9,1.0},
xticklabels={0,0.1,0.2,0.3,0.4,0.5,0.6,0.7,0.8,0.9,1},
ytick={-3,0,3,6,9,12,15},
yticklabels={-3,0,3,6,9,12,15}
]
\addplot [black, mark = *]coordinates {
    (.5 , 13.8567)
    (.6 , 2.9628)
    (.7 , -.1832)
    (.8 , -0.3820)
    (.9 , -0.5531)
    (1  , -0.7059 )
  } ;
\addplot [black, dotted] coordinates {(0,0) (1,0)};
\end{axis}
\end{tikzpicture}
\caption{Variation of $\max\limits_{\boldkappa_1,\boldkappa_2}\left\{D(\boldkappa_1||\tilde{\boldkappa})-0.5D(\boldkappa_2||\tilde{\boldkappa})\right\}$ with $\hat{\lambda}$ for Vector Gaussian distribution.}
\label{fig:Gaussian}
\end{minipage}
\end{figure*}

\subsubsection{Gaussian distribution}
\begin{equation}
f(x;\mu,\sigma^2) = \frac{1}{\sqrt{2\pi \sigma^2}}\exp\left\{-\frac{(x-\mu)^2}{2\sigma^2}\right\}
\end{equation}

We consider two different cases: a) Unknown means and known variance b) Known means and unknown variance. In the latter case, we can subtract the mean value and consider them to be distributions with zero mean.
\paragraph{Unknown means and known variance}
In this case, we have $\boldeta = \frac{\mu}{\sigma}$, $A(\boldeta) = \frac{\boldeta^2}{2}$, $\textbf{T}(x) = \frac{x}{\sigma}$ and $\boldkappa = \frac{\mu}{\sigma}$. We get
\begin{equation}
F(\boldkappa) = \frac{\mu^2}{2\sigma^2},
\end{equation}
and
\begin{equation}
\frac{dF}{d\boldkappa} = \boldkappa, \frac{d^2F}{d\boldkappa^2} = 1.
\end{equation}

This reduces the expression in (\ref{eqn:main_ineq_.5}) to
\begin{equation}
\int\limits_{\hat{\lambda}}^{1} (1-u) du < \frac{1}{2}\int\limits_{0}^{\hat{\lambda}} u du.
\end{equation}
which on solving gives the condition $\hat{\lambda}>0.59$.

\paragraph{Zero mean and unknown variance}
In this case, we have $\boldeta = \frac{-1}{2\sigma^2}$, $\textbf{T}(x) = x^2$, $A(\boldeta) = \log \sigma$, $\boldkappa = \sigma^2$ and
\begin{equation}
F(\boldkappa) = \frac{-1}{2}\left(1+\log \sigma^2\right).
\end{equation}
We obtain
\begin{equation}
\frac{dF}{d\boldkappa} = \frac{-1}{2\boldkappa}, \frac{d^2F}{d\boldkappa^2} = \frac{1}{2\boldkappa^2}
\end{equation}

Since the second derivative is a decreasing function in $u$, we can use the similar analysis as in case of Poisson distribution to obtain bounds on $\hat{\lambda}$ as $\hat{\lambda}>0.59$.

\subsection{Vector parameter distributions}
In this case, we assume both the mean and variances to be unknown.
\begin{equation}
f(x;\mu,\sigma^2) = \frac{1}{\sqrt{2\pi \sigma^2}}\exp\left\{-\frac{(x-\mu)^2}{2\sigma^2}\right\}
\end{equation}
with $\boldeta = \left[\frac{\mu}{\sigma^2} \,\,\, \frac{-1}{2\sigma^2}\right]^T$, $\textbf{T}(x) = \left[x\,\,\, x^2\right]^T$ and $A(\boldeta) = -\frac{\boldeta_1^2}{4\boldeta_2}-\frac{1}{2}\log(-2\boldeta_2)$. The expectation parameter $\boldkappa$ is given as
\begin{equation*}
\boldkappa =
\begin{bmatrix}
\mu \\
\mu^2+\sigma^2
\end{bmatrix}.
\end{equation*}
The dual function $F(\boldkappa)$ is
\begin{equation}
F(\boldkappa) = -\frac{1}{2}-\frac{1}{2}\log(\boldkappa(2)-\boldkappa(1)^2),
\end{equation}
where $\boldkappa(1) = \mu$ and $\boldkappa(2) = \mu^2+\sigma^2$. Computing the Hessian for $F(\cdot)$, we get

\begin{equation}
\nabla^2_{\boldkappa} F = \frac{1}{(\boldkappa(2)-\boldkappa(1)^2)^2}
\begin{bmatrix}
\boldkappa(1)^2+\boldkappa(2) & -\boldkappa(1)\\
-\boldkappa(1) & 1/2
\end{bmatrix}.
\end{equation}
Again, since we do not have an analytical solution for $\hat{\lambda}$ for which (\ref{eqn:main_ineq_.5}) is true, we checked the inequality in Fig. \ref{fig:Gaussian} for $\boldkappa_1$ and $\boldkappa_2$ in the range $[0,20]$ and variances in the range $[1,21]$ for $\hat{\lambda} \in [0,1]$. The search was coarse with $\boldkappa_1, \boldkappa_2$ and variance incremented in steps of 1 unit. Fig. \ref{fig:Gaussian} suggests that the assumption in (\ref{eqn:main_ineq_.5}) may hold for $\hat{\lambda}>0.7$. $\hfill \IEEEQEDopen$

\section{Proofs in the analysis}

\subsection{Proof for finite stopping time (Proposition \ref{finite_stopping_time})} \label{finite_stopping_time_proof}
The proof is carried out in a series of steps. First, we show that the maximum-likelihood estimates of the parameters converge to their true values. We use this result to show that under the non-stopping policy $\tilde{\pi}_{SM}$, the test statistic associated with the index of the odd arm drifts to infinity. This assures that the statistic crosses the threshold in finite time and that the policy stops.

In the proof, we use $\bf{0}$ and $\bf{1}$ to denote the all-zero and all-ones vectors, respectively.
{{
\begin{proposition}\label{ML_conv}
Fix $K\geq 3$. Let $\psi=\left(i,\boldeta_1,\boldeta_2\right)$ be the true configuration. Consider the non-stopping policy $\tilde{\pi}_{SM}$. As $n \rightarrow \infty$ the following convergences hold almost surely:
\begin{equation}
\label{eqn:expectationparameterconvergence}
\hat{\boldkappa}_1(i) = \frac{\textbf{Y}_i^n}{N_i^n} \rightarrow \boldkappa_1 \text{, } \hat{\boldkappa}_2(i) = \frac{\textbf{Y}^n-\textbf{Y}_i^n}{n-N_i^n} \rightarrow \boldkappa_2
\end{equation}
and
\begin{equation}\label{eqn:expectationparameterconvergence2}
\hat{\boldkappa}_1(j) = \frac{\textbf{Y}_j^n}{N_i^n} \rightarrow \boldkappa_2 .
\end{equation}

\begin{equation}\label{eqn:canonicalparameterconvergence}
\hat{\boldeta}_1(i):= \boldeta\left(\hat{\boldkappa}_1(i)\right) \rightarrow \boldeta_1 \text{, } \hat{\boldeta}_2(i) = \boldeta\left(\hat{\boldkappa}_2(i)\right) \rightarrow \boldeta_2
\end{equation}
\begin{equation}
\label{eqn:canonicalparameterconvergence2}
\hat{\boldeta}_1(j) = \boldeta\left(\hat{\boldkappa}_1(j)\right) \rightarrow \boldeta_2.
\end{equation}
Also,
\begin{equation}\label{eqn:canonicalparameterconvergence3}
\boldeta_1^*(i) \rightarrow \boldeta_1 \text{, } \boldeta_2^*(i) \rightarrow \boldeta_2
\end{equation}
\begin{equation}\label{eqn:canonicalparameterconvergence4}
\boldeta_1^*(j) \rightarrow \boldeta_2.
\end{equation}
\end{proposition}}}


\begin{IEEEproof}
Let $\mathcal{F}_{l-1}$ denote the $\sigma$ field generated by $\left(\textbf{T(}X^{l-1}\textbf{)},A^{l-1}\right)$. Consider the martingale difference sequence
\begin{equation*}
\textbf{S}_i^n = \textbf{Y}_i^n-N_i^n\boldkappa_1 = \sum\limits_{l=1}^n\left(\textbf{T(}X_l\textbf{)}-\boldkappa_1\right)1_{A_l=i}.
\end{equation*}
{Since the log partition function $A$ is assumed to be twice continuously differentiable wherever $A$ is finite, we have that  $E\left[\left(\textbf{T(}X_l\textbf{)}-\boldkappa_1\right)\left(\textbf{T(}X_l\textbf{)}-\boldkappa_1\right)^T1_{A_l=i}|\mathcal{F}_{l-1}\right]$ to be finite $\forall l$.} Using the result in \cite[Theorem 1.2A]{victor} we have for any $\epsilon>0$, there exists $c_\epsilon >0$ such that
\begin{equation}
\label{eqn:expo-bound}
P\left({\textbf{S}}_i^n {\succ} n \epsilon {\bf{1}}\right) \leq \exp\left(-c_{\epsilon} n\right).
\end{equation}
By the Borel-Cantelli Lemma, (\ref{eqn:expo-bound}) implies
\begin{equation}\label{eqn:BC_conv_result}
\frac{\textbf{S}_i^n}{n} \rightarrow {{\bf{0}}} \text{ a.s.}
\end{equation}
Further, based on Assumption \ref{sampling}, we have
\begin{equation}\label{eqn:N_i_n_positive}
\liminf\limits_{n \rightarrow \infty}  \frac{N_i^n}{n} \geq c_K \text{ a.s.}
\end{equation}
Combining results in (\ref{eqn:BC_conv_result}) and (\ref{eqn:N_i_n_positive}) we get
\begin{equation}
\frac{\textbf{S}_i^n}{N_i^n} \rightarrow {{\bf{0}}} \text{ a.s.},
\end{equation}
or equivalently,
\begin{equation}
\frac{\textbf{Y}_i^n}{N_i^n} \rightarrow \boldkappa_1 \text{ a.s.}
\end{equation}
Following similar steps, convergences of the other $\textbf{S}_j^n/n$, for $j = 2, 3, \ldots, K$, follow and we get
\begin{equation}
\frac{\textbf{Y}_j^n}{N_j^n} \rightarrow \boldkappa_2.
\end{equation}
Further, these results imply that
\begin{equation}
\frac{\left(\textbf{Y}^n-\textbf{Y}_j^n\right)-\sum\limits_{k \neq j}N_k^n (\boldkappa_1 1_{\{ k=i \}} + \boldkappa_2 1_{\{ k \neq i \}} )}{n-N_j^n} \rightarrow {{\bf{0}}} \text{ a.s.},
\end{equation}
and we get
\begin{equation}
\frac{\textbf{Y}^n-\textbf{Y}_i^n}{n-N_i^n} \rightarrow \boldkappa_2 \text{ a.s.}
\end{equation}
Finally, we use the continuity of the mapping ${\boldeta}\left(\cdot\right)$ to prove the assertions in (\ref{eqn:canonicalparameterconvergence}) and (\ref{eqn:canonicalparameterconvergence2}). {{Next to prove (\ref{eqn:canonicalparameterconvergence3}) and (\ref{eqn:canonicalparameterconvergence4}), note that since $\boldeta_1 \in \Psi_1$ and $\boldeta_2 \in \Psi_2$, for sufficiently large $n$, $\hat{\boldeta}_1(i) \in \Psi_1$, $\hat{\boldeta}_2(i)\in \Psi_2$ and $\hat{\boldeta}_1(j) \in \Psi_1$ by virtue of (\ref{eqn:canonicalparameterconvergence})-(\ref{eqn:canonicalparameterconvergence2}). From (\ref{eqn:ml_choice}), for sufficiently large $n$ we have $\boldeta_1^*(i) = \hat{\boldeta}_1(i)$, $\boldeta_2^*(i)=\hat{\boldeta}_2(i)$ and $\hat{\boldeta}_1(j) = \hat{\boldeta}_1(j)$. Hence, (\ref{eqn:canonicalparameterconvergence})-(\ref{eqn:canonicalparameterconvergence2}) implies (\ref{eqn:canonicalparameterconvergence3})-(\ref{eqn:canonicalparameterconvergence4})}}.
\end{IEEEproof}

{{
\begin{lemma}
\label{lem:regularity}
For any $i$, for any compact set $C$,
\[
  \inf_{\boldeta_i \in C}\| \boldeta_i' - \boldeta_i\| \rightarrow \infty ~ \Rightarrow ~ \inf_{\boldeta_i \in C} D(\boldeta_i \parallel \boldeta_i') \rightarrow \infty.
\]
\end{lemma}
\begin{IEEEproof}
See \cite[Lemma ~8]{general:gayathri} for proof.
\end{IEEEproof}
}}
\begin{lemma}\label{positive_drift}
Fix $K \geq 3$. Let $\psi = \left(i,\boldeta_1,\boldeta_2\right)$ be the true configuration. Consider the non-stopping policy $\tilde{\pi}_{SM}$. Then for all $j\neq i$, we have
\begin{equation}
\liminf\limits_{n \rightarrow \infty} \frac{Z_{ij}\left(n\right)}{n} > 0 \text{ a.s.}
\end{equation}
\end{lemma}

{{
\begin{IEEEproof}
We begin with the expression for $Z_{ij}(n)$ from (\ref{eqn:GLR_statistic}). 
\begin{eqnarray}
\lefteqn{\liminf\limits_{n \rightarrow \infty} \frac{Z_{ij}(n)}{n}}\nonumber\\
 &=& \liminf\limits_{n \rightarrow \infty} \Bigg( \frac{1}{n} \Big[2 \log\mathcal{H}\left(\boldtau,n_0\right) - \log \mathcal{H}\left(\textbf{Y}_i^n+\boldtau,N_i^n+n_0\right)-\log \mathcal{H}\left(\textbf{Y}^n-\textbf{Y}_i^n+\boldtau, n-N_i^n+n_0\right) \nonumber\\
                                   & & ~~~ -\boldeta_1^*(j)^T\left(j\right)\textbf{Y}_j^n+N_j^nA\left(\boldeta_1^*(j)\right)-\boldeta_2^*(j)^T\left(\textbf{Y}^n-\textbf{Y}_j^n\right) +\left(n-N_j^n\right)A\left(\boldeta_2^*(j)\right)\Big]\Bigg)\nonumber \\
                                   & &\\
                                   &=& \liminf\limits_{n \rightarrow \infty} \Bigg(\frac{1}{n} \log \int\limits_{\boldeta'_1} \exp\left[(\boldY_i^n+\boldtau)^T\boldeta_1'-(N_i^n+n_0)A(\boldeta_1')\right] d\boldeta_1' \nonumber\\
                                   & & ~~~~~+\frac{1}{n} \log \int\limits_{\boldeta'_2} \exp\left[(\boldY^n-\boldY_i^n+\boldtau)^T\boldeta_2'-(n-N_i^n+n_0)A(\boldeta_2')\right] d\boldeta_2'\nonumber\\
                                   && ~~~~~-\frac{1}{n}\log\exp\Big(\boldeta_1^*(j)^T\textbf{Y}_j^n-N_j^nA\left(\boldeta_1^*(j)\right)+\boldeta_2^*(j)^T\left(\textbf{Y}^n-\textbf{Y}_j^n\right) -\left(n-N_j^n\right)A\left(\boldeta_2^*(j)\right)\Big)\Bigg)\nonumber\\
                                   & & \\
                                    &=& \liminf\limits_{n \rightarrow \infty} \Bigg(\frac{1}{n} \log \int\limits_{\boldeta'_1} \exp\Big(n\left[\frac{N_i^n}{n}\left(\frac{\boldY_i^n}{N_i^n}+\frac{\boldtau}{N_i^n}\right)^T\boldeta_1'-\frac{N_i^n+n_0}{n}A(\boldeta_1')\right]\Big) d\boldeta_1' \nonumber\\
                                   & & ~~~~~+\frac{1}{n} \log \int\limits_{\boldeta'_2} \exp\Bigg(n\left[\frac{n-N_i^n}{n}\left(\frac{\boldY^n-\boldY_i^n}{n-N_i^n}+\frac{\boldtau}{n-N_i^n}\right)^T\boldeta_2'-\frac{n-N_i^n+n_0}{n}A(\boldeta_2')\right]\Bigg) d\boldeta_2'\nonumber\\
                                   && ~~~~~-\frac{1}{n}\log\exp\Big(n\Bigg[\frac{N_j^n}{n}\boldeta_1^*(j)^T\frac{\textbf{Y}_j^n}{N_j^n}-\frac{N_j^n}{n}A\left(\boldeta_1^*(j)\right)\nonumber\\
                                   &&\qquad \qquad \qquad \qquad +\frac{n-N_j^n}{n}\left(\boldeta_2^*(j)^T\frac{\boldY^n-\boldY_j^n}{n-N_j^n} -A\left(\boldeta_2^*(j)\right)\right)\Bigg]\Big)\Bigg).\nonumber\\
                                  & &
\end{eqnarray}
To further simplify, we consider the terms within the exponential. Re-arranging and re-writing few  terms we obtain,
\begin{eqnarray}
&& \frac{N_i^n}{n}\Big(\left({\boldeta}_1'-\boldeta_2^*(j)\right)^T \hat{\boldkappa}_1-A(\boldeta_1')+A(\boldeta_2^*(j))\Big)+\frac{N_i^n}{n}\left(\boldeta_2^*(j)^T\hat{\boldkappa}_1-A(\boldeta_2^*(j))\right)+\boldeta_1'^T \frac{\boldtau}{n} -\frac{n_0}{n}A(\boldeta_1')\nonumber\\
&&+\frac{n-N_i^n-N_j^n}{n}\Big(\left(\boldeta_2'-\boldeta_2^*(j)\right)^T\hat{\boldkappa}_2-A(\boldeta_2')+A(\boldeta_2^*(j))\Big)+\frac{N_j^n}{n}\left(\boldeta_2'^T\hat{\boldkappa}_2-A(\boldeta_2'\right)\nonumber\\
&& +\frac{n-N_i^n-N_j^n}{n}\left(\boldeta_2^*(j)^T\hat{\boldkappa_2}-A(\boldeta_2^*(j))\right)+\boldeta_2'^T \frac{\boldtau}{n} -\frac{n_0}{n}A(\boldeta_2')\nonumber\\
&&-\frac{N_j^n}{n}\left(\boldeta_1^*(j)^T\hat{\boldkappa}_2+A(\boldeta_1^*(j))\right)-\frac{n-N_j^n}{n}\left(\boldeta_2^*(j)^T\frac{Y^n-Y_j^n}{n-N_j^n}-A(\boldeta_2^*(j))\right)
\end{eqnarray}
Hence we get,
\begin{eqnarray}
&&\lefteqn{\liminf\limits_{n \rightarrow \infty} \frac{Z_{ij}(n)}{n}}\nonumber\\
&=& \liminf\limits_{n \rightarrow \infty} \Bigg\{\nonumber\\
& & \frac{1}{n}\log\int\limits_{\boldeta_1'}\exp\left(n\left[ \frac{N_i^n}{n}\Big(\left({\boldeta}_1'-\boldeta_2^*(j)\right)^T \hat{\boldkappa}_1-A(\boldeta_1')+A(\boldeta_2^*(j))+\boldeta_1'^T \frac{\boldtau}{N_i^n}\Big) -\frac{n_0}{n}A(\boldeta_1')\right]\right)d\boldeta_1'\nonumber\\
&& +\frac{1}{n}\log\int\limits_{\boldeta_2'}\exp\Bigg(n\Bigg[\frac{n-N_i^n-N_j^n}{n}\Bigg(\left(\boldeta_2'-\boldeta_2^*(j)\right)^T\hat{\boldkappa}_2-A(\boldeta_2')+A(\boldeta_2^*(j)) \Bigg)\nonumber\\
&& \qquad \qquad \qquad+\boldeta_2'^T \frac{\boldtau}{n}-\frac{n_0}{n}A(\boldeta_2')+\frac{N_j^n}{n}\left(\left(\boldeta_2'-\boldeta_1^*(j)\right)^T\hat{\boldkappa}_2-A(\boldeta_2')+A(\boldeta_1^*(j))\right)\Bigg)\Bigg]d\boldeta_2'\Bigg\}\nonumber\\
&=& \liminf\limits_{n \rightarrow \infty} \Bigg\{\frac{1}{n}\log\int\limits_{\boldeta_1'}\exp\Bigg(n\Big[ \frac{N_i^n}{n}\Big(\left({\boldeta}_1'-\boldeta_2^*(j)\right)^T {\boldkappa}_1-A(\boldeta_1')+A(\boldeta_2^*(j))\Big)\nonumber\\
& & \qquad \qquad \qquad \qquad +\frac{N_i^n}{n}\left[\left({\boldeta}_1'-\boldeta_2^*(j)\right)^T\left(\hat{\boldkappa}_1- {\boldkappa}_1\right)+ \boldeta_1'^T \frac{\boldtau}{N_i^n}\right] -\frac{n_0}{n}A(\boldeta_1')\Big]\Bigg)d\boldeta_1'\nonumber\\
&& \quad +\frac{1}{n}\log\int\limits_{\boldeta_2'}\exp\Bigg(n\Bigg[\frac{n-N_i^n-N_j^n}{n}\left(\left(\boldeta_2'-\boldeta_2^*(j)\right)^T{\boldkappa}_2-A(\boldeta_2')+A(\boldeta_2^*(j))\right)\nonumber\\
&& \qquad \qquad \qquad +\frac{n-N_i^n-N_j^n}{n}\left[\left(\boldeta_2'-\boldeta_2^*(j)\right)^T\left(\hat{\boldkappa}_2-\boldkappa_2\right)+\boldeta_2'^T \frac{\boldtau}{n-N_i^n-N_j^n}\right]-\frac{n_0}{n}A(\boldeta_2')\nonumber\\
&& \qquad \qquad +\frac{N_j^n}{n}\left(\left(\boldeta_2'-\boldeta_1^*(j)\right)^T{\boldkappa}_2-A(\boldeta_2')+A(\boldeta_1^*(j))+(\boldeta_2'-\boldeta_1^*(j))^T(\hat{\boldkappa}_2-\boldkappa_2)\right)\Bigg]\Bigg)d\boldeta_2'\Bigg\}.\nonumber\\
& & \label{eqn:Zij}
\end{eqnarray}
Note that $\boldeta_i(\boldkappa_i)$ optimises the function $\boldeta_i' \mapsto \boldeta_i'^T\boldkappa_i -A(\boldeta_i')$ for $i=1,2$. We now use this.

Define a ball $\mathcal{B}_\delta(\boldeta_1(\boldkappa_1))$ as an open Euclidean ball of radius $\delta$ around $\boldeta_1(\boldkappa_1)$. Fix $\epsilon_1 >0$. There is then a $\delta>0$ and a $C_\delta >0$ such that, almost surely, for sufficiently large $n$ and for all $\boldeta_i' \in \mathcal{B}_\delta(\boldeta_1(\boldkappa_1))$, we have
\begin{eqnarray*}
||\boldkappa_1-\hat{\boldkappa}_1||_{\infty} & \leq& \epsilon \nonumber\\
\left|(\boldeta_1'-\boldeta^*_2(j))^T(\hat{\boldkappa}_1-\boldkappa_1)+\boldeta_1'^T\frac{\boldtau}{N_i^n}\right| &\leq& C_\delta \epsilon\\
\left|\frac{n_0}{n}A(\boldeta_1')\right| & \leq& C_\delta\\
\left|\boldeta_1'^T\boldkappa_1-\boldeta_1^T\boldkappa_1-A(\boldeta_1')+A(\boldeta_1)\right| &\leq& \epsilon.
\end{eqnarray*}
The second inequality follows from the results of Lemma \ref{lem:regularity} and the fact that the function $\lambda(i)D(\boldeta_1||\tilde{\boldeta})+(1-\lambda(i))\frac{K-2}{K-1}D(\boldeta_2||\tilde{\boldeta})$ is continuous in $\boldeta' \in B_\delta(\boldeta_1)$ and hence bounded.
In a similar way we also have
\begin{eqnarray*}
||\boldkappa_2-\hat{\boldkappa}_2||_{\infty} & \leq& \epsilon_2 \nonumber\\
\left|(\boldeta_2'-\boldeta^*_2(j))^T(\hat{\boldkappa}_2-\boldkappa_2)+\boldeta_2'^T\frac{\boldtau}{n-N_i^n-N_j^n}\right| &\leq& C_\delta \epsilon_2\\
\left|\frac{n_0}{n}A(\boldeta_2')\right| & \leq& C_\delta\\
\left|\boldeta_2'^T\boldkappa_2-\boldeta_2^T\boldkappa_2-A(\boldeta_2')+A(\boldeta_2)\right| &\leq& \epsilon_2\\
\left|(\boldeta_2'-\boldeta_1^*(j))^T(\hat{\boldkappa}_2-\boldkappa_2)\right|\leq C_\delta \epsilon_2\\
\end{eqnarray*}
Further we can lower bound the integral in (\ref{eqn:Zij}) by restricting the integral to the set $B_r(\boldeta(\boldkappa))$. Putting all these together we get
\allowdisplaybreaks{
\begin{align}
\lefteqn{\liminf\limits_{n \rightarrow \infty}\frac{Z_{ij}(n)}{n}} \nonumber\\
& \geq \liminf\limits_{n \rightarrow \infty} \Bigg\{\frac{1}{n}\log\int\limits_{\boldeta_1' \in B_\delta(\boldeta_1)}\exp\Bigg(n\Bigg[ \frac{N_i^n}{n}\Big(\left({\boldeta}_1'-\boldeta_2^*(j)\right)^T {\boldkappa}_1-A(\boldeta_1')+A(\boldeta_2^*(j))\Big)\nonumber\\
&  \qquad \qquad \qquad \qquad +\frac{N_i^n}{n}\left[\left({\boldeta}_1'-\boldeta_2^*(j)\right)^T\left(\hat{\boldkappa}_1- {\boldkappa}_1\right)+ \boldeta_1'^T \frac{\boldtau}{N_i^n}\right] -\frac{n_0}{n}A(\boldeta_1')\Bigg]\Bigg)d\boldeta_1'\nonumber\\
& \qquad \qquad +\frac{1}{n}\log\int\limits_{\boldeta_2'\in B_\delta(\boldeta_2)}\exp\Bigg(n\Bigg[\frac{n-N_i^n-N_j^n}{n}\left(\left(\boldeta_2'-\boldeta_2^*(j)\right)^T{\boldkappa}_2-A(\boldeta_2')+A(\boldeta_2^*(j))\right)\nonumber\\
& \qquad \qquad \qquad +\frac{n-N_i^n-N_j^n}{n}\left[\left(\boldeta_2'-\boldeta_2^*(j)\right)^T\left(\hat{\boldkappa}_2-\boldkappa_2\right)+\boldeta_2'^T \frac{\boldtau}{n-N_i^n-N_j^n}\right]-\frac{n_0}{n}A(\boldeta_2')\nonumber\\
& \qquad \qquad \qquad +\frac{N_j^n}{n}\left(\left(\boldeta_2'-\boldeta_1^*(j)\right)^T{\boldkappa}_2-A(\boldeta_2')+A(\boldeta_1^*(j))+(\boldeta_2'-\boldeta_1^*(j))^T(\hat{\boldkappa}_2-\boldkappa_2)\right)\Bigg]\Bigg)d\boldeta_2'\Bigg\}\nonumber\\
&\\
&\geq  \liminf\limits_{n \rightarrow \infty}\Bigg\{ \frac{1}{n} \log \int\limits_{\boldeta_1' \in B_\delta(\boldeta_1)}\exp\Bigg(n \Bigg[\frac{N_i^n}{n}\Big(\left({\boldeta}_1-\boldeta_2^*(j)\right)^T {\boldkappa}_1-A(\boldeta_1)+A(\boldeta_2^*(j))\Big)\nonumber\\
&  \qquad \qquad \qquad \qquad +\frac{N_i^n}{n}\left(-\epsilon+ (-C_\delta\epsilon)\right)-\frac{C_\delta}{n}\Bigg]\Bigg)d\boldeta_1'\\
& \qquad \qquad  +\frac{1}{n}\log\int\limits_{\boldeta_2'\in B_\delta(\boldeta_2)}\exp\Bigg(n\Bigg[\frac{n-N_i^n-N_j^n}{n}\left(\left(\boldeta_2-\boldeta_2^*(j)\right)^T{\boldkappa}_2-A(\boldeta_2)+A(\boldeta_2^*(j))\right)\nonumber\\
& \qquad \qquad \qquad +\frac{n-N_i^n-N_j^n}{n}\left(-\epsilon+(-C_\delta\epsilon)\right)-\frac{C_\delta}{n}\nonumber\\
& \qquad \qquad \qquad+\frac{N_j^n}{n}\left(\left(\boldeta_2-\boldeta_1^*(j)\right)^T{\boldkappa}_2-A(\boldeta_2)+A(\boldeta_1^*(j))+(-\epsilon-C_\delta\epsilon\right)\Bigg]\Bigg)d\boldeta_2'\Bigg\}\nonumber\\
& \\
&\geq \liminf\limits_{n \rightarrow \infty}\Bigg\{ \frac{N_i^n}{n}D(\boldeta_1||\boldeta_2^*(j)) + \frac{n-N_i^n-N_j^n}{n}D(\boldeta_2||\boldeta_2^*(j))+\frac{N_j^n}{n}D(\boldeta_2||\boldeta_1^*(j))\Bigg\}\nonumber\\
&  +\liminf\limits_{n \rightarrow \infty} \frac{1}{n}\log \left(\text{Leb}(\boldeta_1' \in B_\delta(\boldeta_1))\right) + \liminf\limits_{n \rightarrow \infty} \frac{1}{n}\log \left(\text{Leb}(\boldeta_2' \in B_\delta(\boldeta_2))\right) \nonumber\\
& - \limsup\limits_{n \rightarrow \infty} \left((1+C_\delta)\epsilon + \frac{C_\delta}{n}\right)\\
&\geq \liminf\limits_{n \rightarrow \infty}\Bigg\{ \frac{N_i^n}{n}D(\boldeta_1||\boldeta_2^*(j)) + \frac{n-N_i^n-N_j^n}{n}D(\boldeta_2||\boldeta_2^*(j))+\frac{N_j^n}{n}D(\boldeta_2||\boldeta_1^*(j))-(1+C_\delta)\epsilon)\Bigg\}\label{eqn:leb}\\
&> 0.
\end{align}}
The inequality in (\ref{eqn:leb}) holds since the Lebesgue measure $\text{Leb}(\boldeta_1' \in B_\delta(\boldeta_1))$ is positive. Based on Assumption \ref{sampling} and suitable choice of $\epsilon$, we arrive at the last inequality.
%
%
\end{IEEEproof}}}

\vspace{.3in}
\noindent [Proof of Proposition \ref{finite_stopping_time}]\begin{IEEEproof}
The following inequalities hold almost surely,
\begin{eqnarray}
\tau\left(\pi_{SM}\left(L,\gamma\right)\right) &\leq& \tau\left(\pi_{SM}^i(L,\gamma)\right)\label{eqn:fst_1}\\
										&=& \inf \left\{n \geq 1| Z_i(n) > \log\left(\left(K-1\right){L}\right)\right\} \nonumber\\
										& \leq & \inf \left\{n \geq 1| Z_{ij}(n') > \log\left(\left(K-1\right){L}\right) \forall n' \geq n, \forall j \neq i\right\} \nonumber \\
										& < & \infty, \label{eqn:fst_4}
\end{eqnarray}
where inequality in (\ref{eqn:fst_1}) follows from the definition of the policy $\pi_{SM}^i(L,\gamma)$ and the last inquality follows from the result in Lemma \ref{positive_drift}.
\end{IEEEproof}

\subsection{Proof for upper bound (Proposition \ref{upper_bound})}\label{upper_bound_proof}
The proof is completed in a series of steps. 
%

We begin by showing in Proposition \ref{odd_conv} below that the odd arm chosen by the policy is indeed the odd one. In addition, we also show that the parameters chosen by the policy converge to the true/actual parameters.

\begin{proposition}\label{odd_conv}
Fix $K \geq 3$. Let $\psi=\left(i,\boldeta_1,\boldeta_2\right)$ be the true configuration. Consider the non-stopping policy $\tilde{\pi}_{SM}$. Then as $n \rightarrow \infty$, the following convergences hold almost surely:
\begin{equation}
i^*\left(n\right) \rightarrow i,
\end{equation}
\begin{equation}
\hat{\boldkappa}^n_{1}\left(i^*\left(n\right)\right) \rightarrow \boldkappa_1, \,\,\,
\hat{\boldkappa}^n_{2}\left(i^*\left(n\right)\right) \rightarrow \boldkappa_2,
\end{equation}
\begin{equation}
\hat{\boldeta}^n_{1}\left(i^*\left(n\right)\right) \rightarrow \boldeta_1, \,\,\,
\hat{\boldeta}^n_{2}\left(i^*\left(n\right)\right) \rightarrow \boldeta_2,
\end{equation}
{{
\begin{equation}
{\boldeta}^*_{1}\left(i^*\left(n\right)\right) \rightarrow \boldeta_1, \,\,\,
{\boldeta}^*_{2}\left(i^*\left(n\right)\right) \rightarrow \boldeta_2,
\end{equation}}}
\begin{equation}
\lambda^*\left(i^*\left(n\right),\hat{\boldeta}^n_{1}\left(i^*\left(n\right)\right),\hat{\boldeta}^n_{2}\left(i^*\left(n\right)\right)\right) \rightarrow \lambda^*\left(i,\boldeta_1,\boldeta_2\right),
\end{equation}
{{
\begin{equation}\label{eqn:conv_l_Ni_a}
\frac{N_j^{n,a}}{n} \rightarrow \lambda^*\left(i,\boldeta_1,\boldeta_2\right)(j) \text{ for all } j = 1,2, \ldots, K,
\end{equation}}}
\begin{equation}\label{eqn:conv_l_Ni}
\frac{N_j^n}{n} \rightarrow \lambda^*\left(i,\boldeta_1,\boldeta_2\right)(j) \text{ for all } j = 1,2, \ldots, K,
\end{equation}
\begin{equation}
\frac{\textbf{Y}^n-\textbf{Y}_j^n}{n-N_j^n} \rightarrow \tilde{\boldkappa}\left(\lambda^*\left(i,\boldkappa_1,\boldkappa_2\right)\left(i\right)\right)\text{ for all } j\neq i,
\end{equation}
{{
\begin{equation}\label{eqn:tilde_eta}
\boldeta\left(\frac{Y^n-Y_j^n}{n-N_j^n}\right) \rightarrow \tilde{\boldeta}(\tilde{\boldkappa}),
\end{equation}}}
{{
\begin{equation}\label{eqn:Zi_D*}
\liminf\limits_{n \rightarrow \infty} \frac{Z_i(n)}{n} \geq D^*(i,\boldeta_1,\boldeta_2).
\end{equation}}}
where $\tilde{\boldkappa}$ is as in (\ref{kappa}).
\end{proposition}


\begin{IEEEproof}
The proof is based on the continuity of $\lambda^*$, martingale convergence arguments and the results from Lemma \ref{positive_drift}. For further details refer to \cite[Prop.~12,~p.~21]{vaidhiyan15}. Results for $\boldeta$ follow from the continuity of the function {$\boldeta(\cdot)$ in (\ref{dual})}. 

\vspace{.3in}
{{\noindent Proof of (\ref{eqn:conv_l_Ni}): Let $\{V_1,V_2, \ldots, V_{n^a}\}$ be such that $V_k$ is the number of sluggish instants plus one active instance corresponding to the $k$th active instance, $k = 1,2, \ldots, n^a$. Then $V_t$'s are independent and identical random variables with the geometric distribution of parameter $\gamma$. Additionally, to make the total of $n$ arm pulls at time instant $n$, the last `sluggish run' should also be accounted. We do this by re-writing the expression in (\ref{eqn:num-samples}) as
\begin{equation}
N_i^n = \sum\limits_{t=1}^{n^a} V_t1_{\{A_t=i\}} + \overline{V}_i
\end{equation}
where $\overline{V}_i$ is nonzero for at most for one $i$ and corresponds to the latest sluggish run at time instant $n$. To study the limit of $N_i^n/n$, it suffices to study
\begin{equation}\label{eqn:split_frac}
\frac{1}{n} \sum\limits_{t=1}^{n^a} V_t1_{\{A_t=i\}} = \frac{n^a}{n} \cdot \frac{N_i^{n,a}}{n^a} \cdot \frac{1}{N_i^{n,a}}\sum\limits_{t=1}^{n^a} V_t1_{\{A_t=i\}}.
\end{equation}
We consider each term on the right-hand side of (\ref{eqn:split_frac}) in detail. Note that $n^a/n \rightarrow \gamma$ and from (\ref{eqn:conv_l_Ni_a}) we get $N_i^{n,a}/n^a \rightarrow \lambda^*(i,\boldeta_1,\boldeta_2)$. {{Based on Assumption \ref{sampling} and the fact that the switching parameter $\gamma>0$, we have $N_i^{n,a} \rightarrow \infty$ as $n \rightarrow \infty$.}}  Note that the summation in (\ref{eqn:split_frac}) has $N_i^{n,a}$ terms, and hence the sample mean converges to the expected value of $V_t$ which is $1/\gamma$. Hence,we get, almost surely,
\begin{equation}\label{eqn:exp_frac}
\lim\limits_{n \rightarrow \infty} \frac{N_i^n}{n} = \gamma \cdot \lambda_i^*(\overline{\boldeta}) \cdot \frac{1}{\gamma} = \lambda_i^*(\overline{\boldeta}).
\end{equation}
This concludes the proof of (\ref{eqn:conv_l_Ni}). 

\vspace{.3in}
\noindent Proof of (\ref{eqn:Zi_D*}): Using results from Lemma \ref{positive_drift} and convergence results in (\ref{eqn:conv_l_Ni}) and (\ref{eqn:tilde_eta}) we have
\begin{eqnarray}
\lefteqn{\liminf\limits_{n \rightarrow \infty} \frac{Z_{i}(n)}{n}}\nonumber\\
 &\geq& \liminf\limits_{n \rightarrow \infty}\Bigg\{ \frac{N_i^n}{n}D(\boldeta_1||\boldeta_2^*(j)) + \frac{n-N_i^n-N_j^n}{n}D(\boldeta_2||\boldeta_2^*(j))+\frac{N_j^n}{n}D(\boldeta_2||\boldeta_1^*(j))-(1+C_\delta)\epsilon)\Bigg\}\nonumber\\
&\geq& \lambda_i^*D(\boldeta_1||\tilde{\boldeta})+(1-\lambda_i^*)\frac{K-2}{K-1}D(\boldeta_2||\tilde{\boldeta})\\
&=& D^*(i,\boldeta_1,\boldeta_2).
\end{eqnarray}

}}
\end{IEEEproof}


\begin{lemma}\label{tau20}
Fix $K \geq 3$. Let $\psi = \left(i,\boldeta_1,\boldeta_2\right)$ be the true configuration. Consider the policy $\pi_{SM}\left(L,\gamma\right)$. Then,
\begin{equation}
\liminf\limits_{L \rightarrow \infty}\tau\left(\pi_{SM}\left(L,\gamma\right)\right) \rightarrow \infty \text{ a.s. }
\end{equation}
\end{lemma}
\begin{IEEEproof}
It suffices to show that, as $L \rightarrow \infty$,
\begin{equation}
P\left(\tau\left(\pi_{SM}\left(L,\gamma\right)\right)<n\right) \rightarrow 0 \text{ for all } n.
\end{equation}
We begin with

\begin{eqnarray}
\lefteqn{\limsup\limits_{L \rightarrow \infty} P\left(\tau\left(\pi_{SM}\left(L,\gamma\right)\right)<n\right)}\nonumber\\
&=& \limsup\limits_{L \rightarrow \infty} P \left(\max\limits_{1 \leq l \leq n} Z_j\left(l\right) >\log\left(\left(K-1\right)L\right) \text{ for some } j \right)\nonumber\\
							& \leq & \limsup\limits_{L \rightarrow \infty} \sum\limits_{j=1}^K \sum\limits_{l=1}^n P(Z_j\left(l\right) >\log(\left(K-1\right)L)) \label{tau20_2}\\
							& \leq & \limsup\limits_{L \rightarrow \infty} \frac{1}{\log\left(\left(K-1\right)L\right)} \sum\limits_{j=1}^K \sum\limits_{l=1}^n E\left[N_j^lD\left(\hat{\boldkappa}_1(j)||\boldkappa_0\right)+(l-N_j^l)D\left(\hat{\boldkappa}_2(j)||\boldkappa_0\right)\right]\nonumber\\
							& & \label{tau20_3}\\
							&=& \limsup\limits_{L \rightarrow \infty} \frac{1}{\log\left(\left(K-1\right)L\right)} \sum\limits_{j=1}^K \sum\limits_{l=1}^n	\Big\{l\boldkappa_0^T\boldeta_0-N_j^l\boldeta_0^TE\left[\hat{\boldkappa}_1(j)\right]+N_j^lE\left[F(\hat{\boldkappa}_1(j))\right]\nonumber\\
							& & \qquad \qquad -lF(\boldkappa_0)-(l-N_j^l)\boldeta_0^TE\left[\hat{\boldkappa}_2(j)\right]+(l-N_j^l)E\left[F(\hat{\boldkappa}_2(j)\right]\Big\}\label{tau20_4}\\							
							& \leq & \limsup\limits_{L \rightarrow \infty} \frac{1}{\log\left(\left(K-1\right)L\right)} \sum\limits_{j=1}^K \sum\limits_{l=1}^n \Bigg\{N_j^l\big\{E\left[\hat{\boldkappa}_1^T\left(j\right)\hat{\boldeta}_1(j)\right]-A\left(\boldeta_1\left(j\right)\right)\big\}\nonumber\\
							& &  \qquad \qquad  \qquad \qquad +(l-N_j^l)\left\{E\left[\hat{\boldkappa}_2^T(j)\hat{\boldeta}_2(j)\right]-A\left(\boldeta_2\left(j\right)\right)\right\}\Bigg\} \label{tau20_5} \\
							& = & 0. \label{tau20_6}
\end{eqnarray}
Inequality in (\ref{tau20_2}) follows from union bound. We will demonstrate (\ref{tau20_3}) shortly. Using the expression for $D(\cdot||\cdot)$ from (\ref{eqn:relativeentropy-expectation}) and simplifying we obtain the equality in (\ref{tau20_4}). In inequality (\ref{tau20_5}), we have used the result from \cite[Th 3.1, p.2]{birge} to get an upper bound on $E\left[F(\cdot)\right]$. To obtain (\ref{tau20_6}), we have then used the fact that the expectations are finite.

The inequality in (\ref{tau20_3}), the inequality we are yet to show, is obtained using Markov inequality and the result
\begin{eqnarray}
Z_j\left(l\right) &=& \log \left(\frac{\tilde{f}\left(X^l,A^l|H=j\right)}{\max\limits_{k \neq j}\hat{f}\left(X^l,A^l|H=k\right)}\right) \nonumber\\
	   &\leq & \log \left(\frac{\hat{f}\left(X^l,A^l|H=j\right)}{\hat{f}\left(X^l,A^l|H=k\right)}\right) \text{ for some } k\neq j \nonumber\\
       &=& N_j^lF\left(\hat{\boldkappa}_1\left(j\right)\right)+\left(l-N_j^l\right)F\left(\hat{\boldkappa}_2\left(j\right)\right)-N_k^l F\left(\hat{\boldkappa}_1\left(k\right)\right)-\left(l-N_k^l\right)F\left(\hat{\boldkappa}_2\left(k\right)\right)
	    \label{tau20_9} \\
	   &=& N_j^lD\left(\hat{\boldkappa}_1(j)||\boldkappa_0\right)+(l-N_j^l)D\left(\hat{\boldkappa}_2(j)||\boldkappa_0\right)-l\boldeta_0^T{\boldkappa_0} +\boldeta_0^T \textbf{Y}^l + lF(\boldkappa_0)\nonumber\\
	   & & -\left[N_k^lD\left(\hat{\boldkappa}_1(k)||\boldkappa_0\right)+(l-N_k^l)D\left(\hat{\boldkappa}_2(k)||\boldkappa_0\right)-l\boldeta_0^T{\boldkappa_0} +\boldeta_{0}^T \textbf{Y}^l + lF(\boldkappa_0)\right]\label{tau20_10}\\
	   &\leq&  N_j^lD\left(\hat{\boldkappa}_1(j)||\boldkappa_0\right)+(l-N_j^l)D\left(\hat{\boldkappa}_2(j)||\boldkappa_0\right).\label{tau20_11}
\end{eqnarray}
The equality in (\ref{tau20_9}) is obtained using (\ref{conv_conj}) and (\ref{MLf}). The equality in (\ref{tau20_10}) is obtained by introducing the dual pair $\boldkappa_0$ and $\boldeta_0$, by re-writing (\ref{tau20_9}) in terms of the KL divergence, and by using (\ref{eqn:kappa-update}). To obtain (\ref{tau20_11}), we cancel like terms in (\ref{tau20_10}) and recognise that the KL divergence terms within square brackets therein are nonnegative. This finishes the proof of the lemma.
\end{IEEEproof}

{{\begin{lemma}
Fix $K\geq 3$. Let $\psi = \left(i,\boldeta_1,\boldeta_2\right)$ be the true configuration. Consider the policy $\pi_{SM}(L,\gamma)$. We then have
\begin{equation}\label{eqn:drift_pi_SM}
\liminf\limits_{L \rightarrow \infty} \frac{Z_i\left(\tau\left(\pi_{SM}\left(L,\gamma\right)\right)\right)}{\tau\left(\pi_{SM}\left(L,\gamma\right)\right)} \geq D^*\left(i,\boldeta_1,\boldeta_2\right) \text{ a.s.}
\end{equation}
\end{lemma}
\begin{IEEEproof}
This follows easily from Proposition \ref{odd_conv} and Lemma \ref{tau20}.
\end{IEEEproof}}}

With all the ingredients at hand, we begin the proof for Proposition \ref{upper_bound}.

\vspace*{.1in}

\begin{IEEEproof}[Proof of Proposition \ref{upper_bound}]
There are three main results in Proposition \ref{upper_bound}. We discuss the proofs for each of them in detail.

\begin{IEEEproof}[1. Proof of result in (\ref{eqn:upper_bound1})]
Using the definition of $\tau(\pi_{SM}(L,\gamma))$, we have $Z_i(\tau(\pi_{SM}(L,\gamma)-1)<\log((K-1)L)$ at the previous slot. Using this we get,
\begin{equation}\label{eqn:upper_bound11}
\limsup \limits_{L \rightarrow \infty} \frac{Z_i(\tau(\pi_{SM}(L,\gamma))-1)}{\log(L)} \leq \limsup \limits_{L \rightarrow \infty} \frac{\log((K-1)L)}{\log(L)} = 1.
\end{equation}
Substituting (\ref{eqn:drift_pi_SM}) in (\ref{eqn:upper_bound11}), we get
\begin{eqnarray*}
\limsup\limits_{L \rightarrow \infty} \frac{\tau(\pi_{SM}(L,\gamma))}{\log(L)} &=& \limsup\limits_{L \rightarrow \infty}\frac{\tau(\pi_{SM}(L,\gamma))-1}{\log(L)}\\
					& \leq & \frac{1}{D^*(i,\boldeta_1,\boldeta_2)} \text{ a.s.}
\end{eqnarray*}
\end{IEEEproof}

\begin{IEEEproof}[2. Proof of result in (\ref{eqn:upper_bound2})]
A sufficient condition to establish the convergence of expected stopping time is to show that
\begin{equation}
\limsup\limits_{L \rightarrow \infty} E\left[\exp\left(\frac{\tau(\pi_{SM}(L,\gamma))}{\log(L)}\right)\right] < \infty.
\end{equation}

Let $\epsilon >0$ be an arbitrary constant. {{Define
\begin{equation}\label{eqn:u(L)}
u\left(L\right):=\exp\left(\frac{2\log\left(\left(K-1\right)L\right)}{D^*(i,\boldeta_1,\boldeta_2)\log\left(L\right)}+\frac{1}{\log\left(L\right)}\right).
\end{equation}}}
We then have
\begin{eqnarray}
\lefteqn{\limsup\limits_{L \rightarrow \infty} E\left[\exp\left(\frac{\tau(\pi_{SM}(L,\gamma))}{\log(L)}\right)\right]} \nonumber\\	
&=& \limsup\limits_{L \rightarrow \infty} \int\limits_{x \geq 0} P\left(\frac{\tau(\pi_{SM}(L,\gamma))}{\log(L)} > \log(x)\right) dx\label{eqn:conv_EST1}\\
						&\leq & \limsup\limits_{L \rightarrow \infty} \int\limits_{x \geq 0} P\left(\tau^i(\pi_{SM}(L,\gamma)) > \floor{\log(x)\log(L)}\right) dx	 \label{eqn:conv_EST2} \\
						&\leq & \limsup\limits_{L \rightarrow \infty} \Bigg[u(L) + \int\limits_{x \geq u(L)}P\left(\tau^i(\pi_{SM}(L,\gamma)) > \floor{\log(x)\log(L)}\right)dx\Bigg] \label{eqn:conv_EST3} \\
						& \leq & {{\exp\left(\frac{2}{D^*(i,\boldeta_1,\boldeta_2)}\right)}}\nonumber\\
						& & +\limsup\limits_{L \rightarrow \infty}\sum\limits_{n \geq \floor{\log(u(L))\log(L)}}\exp\left(\frac{n+1}{\log(L)}\right)P\left(\tau^i(\pi_{SM}(L,\gamma))>n\right) \label{eqn:conv_EST4}\\
						& \leq & {{\exp\left(\frac{2}{D^*(i,\boldeta_1,\boldeta_2)}\right)}} \nonumber\\
						& & + \limsup\limits_{L \rightarrow \infty}\sum\limits_{n \geq \floor{\log(u(L))\log(L)}}\exp\left(\frac{n+1}{\log(L)}\right)P\left(Z_i(n)<\log((K-1)L)\right). \label{eqn:conv_EST5}
\end{eqnarray}

The inequality in (\ref{eqn:conv_EST3}) is obtained by upper bounding the integrand probability by 1 for $x < u(L)$. Inequality in (\ref{eqn:conv_EST4}) follows from the fact that $P\left(\tau^i(\pi_{SM}(L,\gamma)) > \floor{\log(x)\log(L)}\right)$ is a constant in the interval
\begin{equation*}
x \in \left[\exp\left(\frac{n}{\log(L)}\right), \exp\left(\frac{n+1}{\log(L)}\right)\right)
\end{equation*}
and that the interval length is upper bounded by $\exp\left(\frac{n+1}{\log(L)}\right)$. To show that the right hand side of (\ref{eqn:conv_EST5}) is finite, it is sufficient to show that
{{\begin{equation}\label{eqn:n}
\text{for all } n \geq \frac{2\log((K-1)L)}{D^*(i,\boldeta_1,\boldeta_2)}
\end{equation}}}
and for sufficiently large $L$, there exists constants $\theta>0$ and $0<B<\infty$ such that
\begin{equation}
P\left(Z_i(n)<\log((K-1)L)\right) < Be^{-\theta n}.
\end{equation}
We next show that such an exponential bound exists.

\begin{lemma}\label{exp_bound}
Fix $K \geq 3$. Fix $L>1$. Let $\psi = \left(i,\boldeta_1,\boldeta_2\right)$ be the true configuration. Let u(L) be as in (\ref{eqn:u(L)}).
Then there exist constant $\theta>0$ and $0<B<\infty$, independent of L, such that for all $n \geq \lfloor \log\left(u\left(L\right)\right)\log\left(L\right)\rfloor$, we have
\begin{equation}\label{eb2}
P\left(Z_i\left(n\right) < \log\left(\left(K-1\right)L\right)\right) < Be^{-\theta n}.
\end{equation}
\end{lemma}

\begin{IEEEproof}
Clearly
\begin{eqnarray*}
P\left(Z_i\left(n\right)<\log\left(\left(K-1\right)L\right)\right)
 &=& P\left(\min\limits_{j \neq i}Z_{ij}\left(n\right)<\log\left(\left(K-1\right)L\right)\right)\\
					  &\leq & \sum\limits_{j \neq i} P\left(Z_{ij}\left(n\right)<\log\left(\left(K-1\right)L\right)\right).
\end{eqnarray*}
It now suffices to show that for every $j \neq i$, the probability term in the above expression is exponentially bounded.

\begin{eqnarray}
\lefteqn{P\left(Z_{ij}\left(n\right) < \log\left(\left(K-1\right)L\right)\right)}\nonumber\\
&\leq& P\Bigg(2\log \Big\{\mathcal{H}\left(\boldtau,n_0\right)\Big\}-\log\Big\{\mathcal{H}\left(\textbf{Y}_i^n+\boldtau,N_i^n+n_0\right)\Big\} -\log\Big\{\mathcal{H}\left(\textbf{Y}^n-\textbf{Y}_i^n+\boldtau, n-N_i^n+n_0\right)\Big\} \nonumber\\
					& & \quad -\hat{\boldeta}_1^T\left(j\right)\textbf{Y}_j^n + N_j^nA\left(\hat{\boldeta}_1\left(j\right)\right)-\hat{\boldeta}_2\left(j\right)^T\left(\textbf{Y}^n-\textbf{Y}_j^n\right) +\left(n-N_j^n\right)A\left(\hat{\boldeta}_2\left(j\right)\right)<\log\left(\left(K-1\right)L\right)\Bigg)\nonumber\\
					& & \label{eb5}
\end{eqnarray}

{{
Re-writing (\ref{eb5}) by adding and subtracting a few terms and using the union bound, we get 
\begin{eqnarray}
\lefteqn{P\left(Z_{ij}\left(n\right)<\log\left(\left(K-1\right)L\right)\right)}\nonumber\\
 &\leq&  P\left(2\log\{\mathcal{H}\left(\boldtau,n_0\right)\}<-\epsilon'n\right) + P\left(-\log\Big\{\mathcal{H}\left(\textbf{Y}_i^n+\boldtau,N_i^n+n_0\right)\Big\}-n\lambda_i^*F\left(\boldkappa_1\right) <-\epsilon'n\right) \nonumber\\
      & & + P\left(-\log\Big\{\mathcal{H}\left(\textbf{Y}^n-\textbf{Y}_i^n+\boldtau, n-N_i^n+n_0\right)\Big\}-n\left(1-\lambda_i^*\right)F\left(\boldkappa_2\right)<-\epsilon'n\right) \nonumber\\
      & & + P\left(N_j^n\left(-\boldeta_1^*(j)^T\frac{Y_j^n}{N_j^n}+A(\boldeta_1^*(j))\right)+n\lambda_j^*F(\boldkappa_2) < -\epsilon'n\right)\nonumber\\
      & & +P\left((n-N_j^n)\left(-\boldeta_2^*(j)^T\frac{Y^n-Y_j^n}{n-N_j^n}+A(\boldeta_2^*(j))\right)+n(1-\lambda_j^*)F(\tilde{\boldkappa})<-\epsilon'n\right) \nonumber\\     
      & & + P\left(-\lambda_i^*\left(\tilde{\boldkappa}-\boldkappa_1\right)^T\tilde{\boldeta}-\left(1-\lambda_i^*\right)\frac{K-2}{K-1}\left(\tilde{\boldkappa}-\boldkappa_2\right)^T\tilde{\boldeta}<-\epsilon'n \right)\nonumber \\
      & & + P\left(nD^*(i,\boldeta_1,\boldeta_2)-6\epsilon'n<\log\left(\left(K-1\right)L\right)\right).  \label{eb6}
\end{eqnarray}
We next obtain a bound for each term in (\ref{eb6}). 

\vspace{.3in}
\noindent (i) We begin with the last term. Let
\begin{equation}
\epsilon = \frac{D^*(i,\boldeta_1,\boldeta_2)}{D^*(i,\boldeta_1,\boldeta_2)-6\epsilon'} -1,
\end{equation}
and 
\begin{equation}
n_1 = 2\frac{\log((K-1)L)}{D^(i,\boldeta_1,\boldeta_2)} > \frac{(1+\epsilon)\log((K-1)L)}{D^*(i,\boldeta_1,\boldeta_2)}.
\end{equation}
We then have for $n>n_1$,
\begin{equation}
n\left(D^*(i,\boldeta_1,\boldeta_2)-6\epsilon'\right) > (1+\epsilon)\frac{\log((K-1)L)}{D^*(i,\boldeta_1,\boldeta_2)}\left[D^*(i,\boldeta_1,\boldeta_2)-6\epsilon'\right] = \log((K-1)L).
\end{equation}
Hence we get for $n>n_1$,
\begin{equation}
P\left(nD^*(i,\boldeta_1,\boldeta_2)-6\epsilon'n < \log((K-1)L)\right) = 0.
\end{equation}

\vspace{.3in}
\noindent (ii) Consider next the first term in (\ref{eb6}):
\begin{equation}
P\left(2\log \mathcal{H}_l(\boldtau,n_0) < -\epsilon'n\right).
\end{equation}
The right-hand side inside the probability goes to negative infinity whereas, the left-hand side is a constant. Hence, the probability of the event under study is zero for all sufficiently large $n$.

\vspace{.3in}
\noindent (iii) Consider the second term in \ref{eb6}:
\begin{eqnarray}
\lefteqn{P\left(-\log\Big\{\mathcal{H}\left(\textbf{Y}_i^n+\boldtau,N_i^n+n_0\right)\Big\}-n\lambda_i^*F\left(\boldkappa_1\right) <-\epsilon'n\right)}\nonumber\\
&\leq& P\left(-\frac{1}{n}\log\Big\{\mathcal{H}\left(\textbf{Y}_i^n+\boldtau,N_i^n+n_0\right)\Big\}-\lambda_i^*F\left(\boldkappa_1\right) <-\epsilon', \left|\frac{N_i^n}{n}-\lambda_i^*\right| \leq \epsilon_1,\left\|\frac{Y_i^n}{N_i^n}-\boldkappa_1\right\|_\infty \leq \epsilon_2\right)\nonumber\\
&& +P\left(\left|\frac{N_i^n}{n}-\lambda_i^*\right| > \epsilon_1\right) + P\left(\left\|\frac{Y_i^n}{N_i^n}-\boldkappa_1\right\|_\infty > \epsilon_2\right)\label{eqn:2_eb6}
\end{eqnarray}
Under the conditions 
\begin{equation}
\left|\frac{N_i^n}{n}-\lambda_i^*\right| \leq \epsilon_1 \text{ and } \left\|\frac{Y_i^n}{N_i^n}-\boldkappa_1\right\|_\infty \leq \epsilon_2,
\end{equation}
we next obtain lower bound to $-\frac{1}{n}\log\left\{\mathcal{H}\left(\textbf{Y}_i^n+\boldtau,N_i^n+n_0\right)\right\}$. 
\begin{eqnarray}
\lefteqn{-\frac{1}{n}\log\Big\{\mathcal{H}\left(\textbf{Y}_i^n+\boldtau,N_i^n+n_0\right)\Big\}}\nonumber\\
&=& \frac{1}{n} \log \int\limits_{\boldeta_1'} \exp\left\{n\left[\frac{N_i^n}{n}\left(\frac{Y_i^n}{N_i^n}+\frac{\boldtau}{N_i^n}\right)^T\boldeta_1'-\frac{N_i^n+n_0}{n}A(\boldeta_1')\right]\right\} d\boldeta_1'\\
&=& \frac{1}{n}\log\int\limits_{\boldeta_1'} \exp\left\{n\left[\frac{N_i^n}{n}\left(\boldeta_1'^T\frac{Y_i^n}{N_i^n}-A(\boldeta_1')\right) + \boldeta_1'^T\frac{\boldtau}{n} -\frac{n_0}{n}A(\boldeta_1')\right]\right\}d\boldeta_1'\label{eqn:lb_H}
\end{eqnarray}

Note that the $\boldeta_i$ optimises the function $\boldeta_i' \mapsto \boldeta_i'^T\boldkappa_i-A(\boldeta_i')$. Fix a $\delta>0$. Almost surely, there is a $C_\delta>0$ such that for sufficiently large $n$, we have
\begin{equation*}
||\boldkappa_i-\hat{\boldkappa}_i||_\infty \leq \epsilon_2, 
\end{equation*}
and further for all $\boldeta_1' \in B_\delta(\boldeta)$ we have:
\begin{equation*}
\left|\left(\frac{N_i^n}{n}-\lambda_i^*\right)\left[\boldeta_1'\boldkappa_1-A(\boldeta_1')\right]\right| \leq C_\delta\epsilon_1
\end{equation*}
\begin{equation*}
\left|\boldeta_1'^T\boldtau - n_0A(\boldeta_1')\right| \leq C_\delta
\end{equation*}
\begin{equation*}
\left|\frac{N_i^n}{n}\boldeta_1'^T(\hat{\boldkappa}_1-\boldkappa_1)\right| \leq C_\delta\epsilon_2
\end{equation*}
\begin{equation*}
\left|\boldeta_1'^T\boldkappa_1-A(\boldeta_1')-\left(\boldeta_1^T\boldkappa_1-A(\boldeta_1)\right)\right| \leq \tau(\delta)
\end{equation*}
where in the last inequality, $\tau(\delta) \rightarrow 0$ as $\delta \rightarrow 0$ due to the continuity of $A(\cdot)$. Putting all these ideas together, we can lower bound the integral in (\ref{eqn:lb_H}):
\begin{eqnarray}
\lefteqn{-\frac{1}{n}\log\Big\{\mathcal{H}\left(\textbf{Y}_i^n+\boldtau,N_i^n+n_0\right)\Big\}}\nonumber\\
&\geq& \frac{1}{n}\log\int\limits_{\boldeta_1' \in B_\delta(\boldeta_1)} \exp\left\{n\left[\frac{N_i^n}{n}\left(\boldeta_1'^T\hat{\boldkappa}_1-A(\boldeta_1')\right)\right]-C_\delta\right\}d\boldeta_1'\\
&=& \frac{1}{n}\log\int\limits_{\boldeta_1' \in B_\delta(\boldeta_1)}\exp\Bigg\{n\lambda_i^*(\boldeta_1'^T\hat{\boldkappa}_1-A(\boldeta_1'))+n\left(\frac{N_i^n}{n}-\lambda_i^*\right)(\boldeta_1'^T\boldkappa_1-A(\boldeta_1'))\nonumber\\
& & \qquad \qquad \qquad \qquad +n \left(\frac{N_i^n}{n}\boldeta_1'^T(\hat{\boldkappa_1-\boldkappa)})\right)-C_\delta\Bigg\}d\boldeta_1'\\
&=& \frac{1}{n}\log\int\limits_{\boldeta_1' \in B_\delta(\boldeta_1)}\exp\Bigg\{n\lambda_i^*F(\boldkappa_1)+n\lambda_i^*\left(\boldeta_1'^T\boldkappa_1-A(\boldeta_1')-F(\boldkappa_1)\right)\nonumber\\
& & \qquad \qquad +n\left(\frac{N_i^n}{n}-\lambda_i^*\right)(\boldeta_1'^T\boldkappa_1-A(\boldeta_1')) +n \left(\frac{N_i^n}{n}\boldeta_1'^T(\hat{\boldkappa_1-\boldkappa)})\right)-C_\delta\Bigg\}d\boldeta_1'\\
&\geq& \lambda_i^*F(\boldkappa_1)+\frac{1}{n}\log\left(\text{Leb}(\boldeta_1' \in B_\delta(\boldeta_1))\right) - \tau(\delta)-C_\delta(\epsilon_1+\epsilon_2)-\frac{C_\delta}{n}
\end{eqnarray}
Hence, we can upper bound the first term in the RHS of (\ref{eqn:2_eb6}) as follows:
\begin{align}
&{P\left(-\frac{1}{n}\log\Big\{\mathcal{H}\left(\textbf{Y}_i^n+\boldtau,N_i^n+n_0\right)\Big\}-\lambda_i^*F\left(\boldkappa_1\right) <-\epsilon', \left|\frac{N_i^n}{n}-\lambda_i^*\right| \leq \epsilon_1,\left\|\frac{Y_i^n}{N_i^n}-\boldkappa_1\right\|_\infty \leq \epsilon_2\right)}\nonumber\\
&\leq P\left(\frac{1}{n}\log\left(\text{Leb}(\boldeta_1' \in B_\delta(\boldeta_1))\right) - \tau(\delta)-C_\delta(\epsilon_1+\epsilon_2)-\frac{C_\delta}{n}<-\epsilon'\right)\label{eqn:1_2_eb6}
\end{align}
We can ensure that the event within the probability on the RHS does not occur for sufficiently large $n$ by suitable choice of $\delta$, $\epsilon_1$ and $\epsilon_2$. Exponential bounds for the remaining terms in RHS of (\ref{eqn:2_eb6}) follows similar steps as that in (\ref{eqn:expo-bound}). Analysis for the third term in (\ref{eb6}) follows similar steps as that for the second term of (\ref{eb6}).

\vspace{.3in}

\noindent (iv) Consider the fourth term in (\ref{eb6}):
\begin{eqnarray}
\lefteqn{P\left(N_j^n\left(-\boldeta_1^*(j)^T\frac{Y_j^n}{N_j^n}+A(\boldeta_1^*(j))\right)+n\lambda_j^*F(\boldkappa_2) < -\epsilon'n\right)}\nonumber\\
&\leq& P\left(\frac{N_j^n}{n}\left(\boldeta_1^*(j)^T\frac{Y_j^n}{N_j^n}-A(\boldeta_1^*(j))\right)-\lambda_j^*F(\boldkappa_2) > \epsilon', \left|\frac{N_j^n}{n}-\lambda^*_j\right| \leq \epsilon_3,\left\|\frac{Y_j^n}{N_j^n}-\boldkappa_2\right\|_\infty \leq \epsilon_4\right)\nonumber\\
& & +P\left(\left|\frac{N_j^n}{n}-\lambda^*_j\right| > \epsilon_3\right) + P\left(\left\|\frac{Y_j^n}{N_j^n}-\boldkappa_2\right\|_\infty > \epsilon_4\right).\label{eqn:4_eb6}
\end{eqnarray}

Under the conditions
\begin{equation}
\left|\frac{N_j^n}{n}-\lambda^*_j\right| \leq \epsilon_3 \text{ and }\left\|\frac{Y_j^n}{N_j^n}-\boldkappa_2\right\|_\infty \leq \epsilon_4,
\end{equation}
we can re-write the RHS of the event within the first probability term in (\ref{eqn:4_eb6}) as:
\begin{eqnarray}
&=&\frac{N_j^n}{n}\left(-D(\hat{\boldeta}_2||\boldeta_1^*(j))+\hat{\boldeta}_2^T\hat{\boldkappa}_2-A(\hat{\boldeta}_2)\right)-\lambda_j^*\left(D(\boldeta_2||\hat{\boldeta}_2)-\hat{\boldeta}_2^T\boldkappa_2+A(\hat{\boldeta}_2)\right)\\
&=& \left(\frac{N_j^n}{n}-\lambda_j^*\right)\left(-D(\hat{\boldeta}_2||\boldeta_1^*(j))+\hat{\boldeta}_2^T\hat{\boldkappa}_2-A(\hat{\boldeta}_2)\right)-\lambda_j^*\left(D(\boldeta_2||\hat{\boldeta}_2)-\hat{\boldeta}_2^T\boldkappa_2+A(\hat{\boldeta}_2)\right)\nonumber\\
& & +\lambda_j^*\left(-D(\hat{\boldeta}_2||\boldeta_1^*(j))+\hat{\boldeta}_2^T\hat{\boldkappa}_2-A(\hat{\boldeta}_2)\right)\\
&=& \left(\frac{N_j^n}{n}-\lambda_j^*\right)\left(-D(\hat{\boldeta}_2||\boldeta_1^*(j))+\hat{\boldeta}_2^T\hat{\boldkappa}_2-A(\hat{\boldeta}_2)\right)-\lambda_j^*\left(D(\boldeta_2||\hat{\boldeta}_2)+D(\hat{\boldeta}_2||\boldeta_1^*(j))\right)\nonumber\\
& & + \lambda_j^*\left(\hat{\boldeta}_2^T\hat{\boldkappa}_2-\hat{\boldeta}_2^T\boldkappa_2\right)
\end{eqnarray}
we can re-write the first term in the RHS of (\ref{eqn:4_eb6}) as
\begin{eqnarray}
\lefteqn{P\left(\frac{N_j^n}{n}\left(-\boldeta_1^*(j)^T\frac{Y_j^n}{N_j^n}+A(\boldeta_1^*(j))\right)+\lambda_j^*F(\boldkappa_2) < -\epsilon', \left|\frac{N_j^n}{n}-\lambda^*_j\right| \leq \epsilon_3,\left\|\frac{Y_j^n}{N_j^n}-\boldkappa_2\right\|_\infty \leq \epsilon_4\right)}\nonumber\\
&=& P\Bigg(\left(\frac{N_j^n}{n}-\lambda_j^*\right)\left(-D(\hat{\boldeta}_2||\boldeta_1^*(j))+\hat{\boldeta}_2^T\hat{\boldkappa}_2-A(\hat{\boldeta}_2)\right)-\lambda_j^*\left(D(\boldeta_2||\hat{\boldeta}_2)+D(\hat{\boldeta}_2||\boldeta_1^*(j))\right)\nonumber\\
& & + \lambda_j^*\left(\hat{\boldeta}_2^T\hat{\boldkappa}_2-\hat{\boldeta}_2^T\boldkappa_2\right), \left|\frac{N_j^n}{n}-\lambda^*_j\right| \leq \epsilon_3,\left\|\frac{Y_j^n}{N_j^n}-\boldkappa_2\right\|_\infty \leq \epsilon_4\Bigg).\label{eqn:1_4_eb6}
\end{eqnarray}

Following steps similar that led to bound LHS of (\ref{eqn:1_2_eb6}) and using the fact that $D(\cdot||\cdot) \geq 0$, we can obtain bounds for (\ref{eqn:1_4_eb6}). Analysis for the fifth term in (\ref{eb6}) follows similar steps as that for the fourth term of (\ref{eb6}).

\vspace{.3in}
\noindent (v) Consider the sixth term in (\ref{eb6}):
\begin{equation}
P\left(-\lambda_i^*\left(\tilde{\boldkappa}-\boldkappa_1\right)^T\tilde{\boldeta}-\left(1-\lambda_i^*\right)\frac{K-2}{K-1}\left(\tilde{\boldkappa}-\boldkappa_2\right)^T\tilde{\boldeta}<-\epsilon'n \right)
\end{equation}
Using the expression for $\tilde{\boldkappa}$ from (\ref{eqn:kappatilde}) and simplifying the LHS of the inquality, we get LHS to be $0$. Hence, we get that the probability of the event is zero.
}}
\end{IEEEproof}

Lemma \ref{exp_bound} finishes the proof for result in (\ref{eqn:upper_bound2}).
\end{IEEEproof}

{\em 3. Proof of (\ref{eqn:upper_bound3})}: To prove this, observe that
\begin{eqnarray}
E[C\left(\pi_{SM}\left(L,\gamma\right)|\psi\right)] &=& E\Big[\tau\left(\pi_{SM}\left(L,\gamma\right)|\psi\right) + \sum\limits_{l=1}^{\tau\left(\pi_{SM}\left(L,\gamma\right)\right)-1} g\left(A_l,A_{l+1}\right)\Big]\nonumber\\
							 &\leq & E[\tau\left(\pi_{SM}\left(L,\gamma\right)|\psi\right)]+ g_{max}E\Big[\sum\limits_{l=1}^{\tau\left(\pi_{SM}\left(L,\gamma\right)\right)-1} 1_{\{A_l \neq A_{l+1}\}}\Big]\nonumber\\
							 &\leq & E[\tau\left(\pi_{SM}\left(L,\gamma\right)|\psi\right)] + g_{max}E\Big[\sum\limits_{l=1}^{\tau\left(\pi_{SM}\left(L,\gamma\right)\right)-1} U_{l+1}\Big]\nonumber\\
							 &=& E[\tau\left(\pi_{SM}\left(L,\gamma\right)|\psi\right)] + g_{max} \gamma E[\tau\left(\pi_{SM}\left(L,\gamma\right)\right)-1]\nonumber\\
							 & \leq & E[\tau\left(\pi_{SM}\left(L,\gamma\right)|\psi\right)]\left(1+g_{max} \gamma\right).\nonumber
\end{eqnarray}
Divide by $\log L$ and let $L \rightarrow \infty$ to get the required result. This completes the proof of (\ref{eqn:upper_bound3}), completes the proof of all three results in the proposition, and thus finishes the proof of Proposition~\ref{upper_bound}.
\end{IEEEproof}

\bibliographystyle{IEEEtran}
\bibliography{ref}

\end{document}